\documentclass[conference]{IEEEtran}
\IEEEoverridecommandlockouts

\usepackage[ruled,linesnumbered]{algorithm2e}

\def\BibTeX{{\rm B\kern-.05em{\sc i\kern-.025em b}\kern-.08em
    T\kern-.1667em\lower.7ex\hbox{E}\kern-.125emX}}
    
\usepackage{CJKutf8}  

\usepackage{pifont}
\usepackage{amssymb}
\usepackage{bbding} 
\usepackage{pifont}
\newcommand{\cmark}{\textcolor{green!80!black}{\ding{51}}}
\newcommand{\xmark}{\textcolor{red}{\ding{55}}}

\usepackage{wrapfig} 

\usepackage{multicol}
\usepackage{multirow}

\usepackage{stackengine}
\usepackage{scalerel}
\usepackage{xcolor}
\newcommand\dangersign[1][2ex]{%
  \renewcommand\stacktype{L}%
  \scaleto{\stackon[1.3pt]{\color{red}$\triangle$}{\tiny !}}{#1}%
}

\usepackage[most]{tcolorbox}

\usepackage{pgfplots}
\usepackage{tikz}
\usepackage{comment}

\usepackage{ulem} 

\usepackage{makecell}
\usepackage{colortbl}

\usepackage{filecontents}
\usepackage{threeparttable}

\usepackage{fancyhdr} 
\usepackage{multirow}
\usepackage{cite}
\usepackage{amsmath,amssymb,amsfonts}

\usepackage{textcomp}
\usepackage{xcolor}

\usepackage{subfigure}
\usepackage{pgfplots}
\usepackage{filecontents}

\usepackage{graphicx,subfigure}

\usepackage{url}
\usepackage{hyperref}
\hypersetup{
    colorlinks=true,
    linkcolor=blue,
    filecolor=magenta,      
    urlcolor=magenta,
    citecolor = magenta,
    pdftitle={An Example},
    pdfpagemode=FullScreen,
    }
\usepackage{url}

\definecolor{bblue}{HTML}{4F81BD}
\definecolor{rred}{HTML}{C0504D}
\definecolor{ggreen}{HTML}{9BBB59}
\definecolor{ppurple}{HTML}{9F4C7C}

\usepackage{longtable}

\usepackage{ulem}

\definecolor{uncommon}{rgb}{0.12, 1, 0}
\definecolor{rare}{rgb}{0, 0.44, 0.87}
\definecolor{epic}{rgb}{0.64, 0.21, 0.93}
\definecolor{legendary}{rgb}{1, 0.5, 0}
\definecolor{mythic}{rgb}{0.9, 0.8, 0.5}

\usepackage{listings}

\usepackage{blindtext}
\usepackage{etoolbox,xstring,mfirstuc,textcase}
\usepackage{booktabs}
\usepackage{pgfplots}

\usepackage{tabularx,booktabs,makecell,multirow, caption}
\usepackage{enumitem} 

\newcolumntype{L}{>{\arraybackslash}X}

\usepackage{xcolor}
\usepackage{tikz}
\usepackage{pgfplots}

\definecolor{findOptimalPartition}{HTML}{D7191C}
\definecolor{storeClusterComponent}{HTML}{FDAE61}
\definecolor{dbscan}{HTML}{ABDDA4}
\definecolor{constructCluster}{HTML}{2B83BA}

\usepackage{etoolbox}
\usepackage{tkz-euclide}
\tikzset{%
    pics/sema/.style args={#1/#2/#3}{code={%
        \ifstrequal{#2}{0}{%
            \node[circle,minimum width=1mm,draw,fill=#1] {};
        }{%
            \tkzDefPoint(0,0){O}
            \tkzDrawSector[R,fill=#1](O,1mm)(90,90-#2)
            \tkzDrawSector[R,fill=#3](O,1mm)(90-#2,90-360)
    }
    }},
}

\usepackage[utf8]{inputenc}
\usepackage{listings}

\begin{document}
\title{BRC-20: Hope or Hype}


\author{\IEEEauthorblockN{Qin Wang\IEEEauthorrefmark{1}, Guangsheng Yu\IEEEauthorrefmark{1}
}

\IEEEauthorrefmark{1} \textit{CSIRO Data61,  Australia}
}

\maketitle

\begin{abstract}
BRC-20 (short for Bitcoin Request for Comment 20) token mania was a key storyline in the middle of 2023. Setting it apart from conventional ERC-20 token standards on Ethereum, BRC-20 introduces non-fungibility to Bitcoin through an editable field in each satoshi (0.00000001 Bitcoin, the smallest unit), making them unique. In this paper, we pioneer the exploration of this concept, covering its intricate mechanisms, features, and state-of-the-art applications. By analyzing the multi-dimensional data spanning over months with factual investigations, we conservatively comment that while BRC-20 expands Bitcoin's functionality and applicability, it may still not match Ethereum's abundance of decentralized applications and similar ecosystems.
\end{abstract}

\vspace{-0.5em}

\section{Introduction}\label{sec-intro}

To date (August 2023), Bitcoin~\cite{nakamoto2019bitcoin} has been operating successfully for 15 years. In terms of market capitalization\footnote{Global ranking, \url{https://companiesmarketcap.com/} [August 2023].}, it currently holds the position of the 10th largest asset globally (US\$590.74b), just behind Berkshire Hathaway (US\$773.17b). Moreover, within the cryptocurrency space, Bitcoin remains the dominant player, accounting for over 53\% of the market share\footnote{Cryptocurrency charts, \url{https://coinmarketcap.com/charts/} [August 2023].}, far surpassing the second-ranking crypto-asset ETH (19.1\%, Ethereum native token~\cite{wood2014ethereum}). Despite its dominance, applications leveraging or operating on Bitcoin have been scarce due to its UTXO data structure~\cite{delgado2019analysis}, limiting its extensibility. Fortunately, recent developments with the emergence of a Bitcoin-fitted standard may change this situation.

BRC-20, or Bitcoin Request for Comment 20~\cite{binancebrc}, is modeled after the Ethereum token standard indexed with ERC-20~\cite{erc20} and was introduced in March 2023 by an anonymous developer known as Domo \cite{brc20experiment}.  BRC-20 is basically Bitcoin’s version of ERC-20, even with some major caveats like a lack of smart contracts. The similar parts come from it being the first token standard defined in Bitcoin, while the key distinction is that BRC-20 incorporates non-fungibility features from ERC-721~\cite{erc20}, making it a hybrid standard encompassing both ERC-20 and ERC-721 functionalities.

In Ethereum, non-fungible tokens (NFTs)~\cite{wang2021non} are implemented through smart contracts, where each user is assigned a unique token ID to claim ownership of a specific asset, such as JPEG files or Crypto punk images, stored off-chain on a server. In contrast, BRC-20 tokens are created through  processes called \textit{ordinal} and \textit{inscription} (cf. Sec.\ref{sec-construction}), which involves adding data to identifiable satoshis (the smallest unit of Bitcoin, 0.00000001 BTC). This data can represent user-customized metadata, ranging from unique identifiers to images, and is stored on-chain. When BRC-20 tokens are transferred, the inscribed data on the satoshis is also transferred via transactions, allowing users to mint NFTs on the Bitcoin network.

BRC-20 has prominently emerged as a focal point within the Bitcoin network, commanding significant attention as underscored by an array of market indicators including Bitcoin's block size, mempool transactions, and transaction fees. During the fervor of the BRC-20 period spanning from early February 2023 to May 2023, several notable developments occurred~\cite{binance1}: (i) The average block size of Bitcoin experienced a substantial surge, leaping from 1.2MB to over 2MB. (ii) The volume of transactions within the memory pool demonstrated a consistent upward trajectory, nearing the 25,000 transaction mark. This contrasts with the relatively stable level of around 5,000 transactions that characterized much of 2022. (iii) Ordinal transaction fees exhibited a steady rise, concurrently driving an approximate 10\% increase in non-Ordinal transaction fees throughout the entirety of March. (iv) The cumulative fees accrued from the minting of Ordinal Inscriptions have now surpassed the 150 BTC milestone. Beyond that, various associated platforms/tools have further contributed to this trend: (v) Statistical resources like Ordinal Wallet~\cite{ordinalwallet}, UniSat~\cite{unisat}, and Dune Analytics~\cite{duneanalytic}\cite{duneanalytic1} also corroborate the upward trajectory in minted Ordinals.

\smallskip
\noindent\textbf{Gaps in user perception.} Despite BRC's remarkable achievements within a short timeframe, its awareness remains surprisingly low. Even among seasoned blockchain researchers and developers (as gathered through informal random surveys without recorded responses), it's evident that very few are acquainted with BRC, Bitcoin token standards, or Bitcoin NFTs. Moreover, our explorations also unveiled that existing resources are inadequate for newcomers. While there are initial introductions to the concept (cf. the \textit{final} paragraph of Sec.\ref{sec-intro}), they largely focus on providing a basic operational overview without digging into the multifaceted aspects involved. This realization motivates our pursuit of understanding this intriguing yet ``enigmatic'' term, and discerning its essence as either a beacon of \textit{hope} or a product of \textit{hype}.

\smallskip
\noindent\textbf{Our attempts.} We approach this via three fundamental pillars.

\smallskip
\noindent\textcolor{teal}{\ding{49} \textit{Systematical contributions}}. We extensively dive into the available open-access resources, encompassing blogs, wikis, forum posts, news articles, Git repositories, and a limited number of scholarly works, based on which, we methodically organize and present a clear and concise understanding of \textit{what BRC is} and \textit{how it functions} (Sec.\ref{sec-construction}), marking a pioneering step in current research. Our exposition commences with an exploration of the fundamental structure of Bitcoin  (Sec.\ref{subsec-const-utxo}) and progresses to elaborate on distinctive aspects like ordinals  (Sec.\ref{subsec-const-ordinal}) and inscriptions (Sec.\ref{subsec-const-inscript}), forming pivotal procedures within the BRC operation.

\smallskip
\noindent\textcolor{teal}{\ding{49} \textit{Quantitative contributions.}} We embark on a comprehensive series of quantitative investigations across multiple dimensions to unveil the genuine dynamics and sentiment prevailing within the market. Our approach involves a meticulous examination of the market performance  (Sec.\ref{sec-investi-token}) of a carefully selected group of representative tokens—comprising three BRC-20 and five ERC-20 projects—spanning a period of four months from the ascent of BRC to the point of composing this study. This analysis encompasses an assessment of various factors including price fluctuations, duration of popularity, market capitalization, and daily transaction volumes. Subsequently, we delve into the user responses evident in social media platforms through tweets  (Sec.\ref{sec-investi-sentiment}) featuring specific hashtags during a randomly chosen recent week. This investigation involves the scrutiny of post content, languages used, influencers contributing to discussions, and the identification of potential fraudulent activities. Additionally, we delve into the historical mainstream prices of tokens (Sec.\ref{sec-historical}), delineating the trajectory of each token wave to ascertain the presence of a potential new BRC-formed wave.

\smallskip
\noindent\textcolor{teal}{\ding{49} \textit{Qualitative contributions.}} We conduct a qualitative exploration  (Sec.\ref{sec-discussion}) that involves juxtaposing BRC-20 against established token standards (Sec.\ref{subsec-quali-compri}). Through this comparison, we derive both the advantages  (Sec.\ref{subsec-quali-advan}) and intrinsic limitations  (Sec.\ref{subsec-quali-disadv}) of BRC-20. Building upon these observations (together with quantitative results), we further compile a review of the actualities and misconceptions present within user perceptions (Sec.\ref{subsec-percep-reality}), culminating in our proposed implications to mitigate these aspects (Sec.\ref{subsec-percep-enhance}).

\begin{table}[!hbt]
\caption{Our results by findings}
\label{tab-result}
\resizebox{\linewidth}{!}{
\begin{tabular}{c|clcccccccc|cc} 

\multicolumn{2}{c}{\textbf{\textit{Findings}}} &  \multicolumn{1}{c}{\textbf{\textit{KeyWords}}} &  \rotatebox{90}{\textbf{\textit{TokenPrice}}} & \rotatebox{90}{\textbf{\textit{PopDuration}}} & \rotatebox{90}{\textbf{\textit{MarketCap}}} & \rotatebox{90}{\textbf{\textit{UserScale}}} & \rotatebox{90}{\textbf{\textit{Daily Tx}}} & \rotatebox{90}{\textbf{\textit{Tweets}}} & \rotatebox{90}{\textbf{\textit{Designs}}} & \rotatebox{90}{\textbf{\textit{Applications}}} & \rotatebox{90}{\textbf{\textit{Hope}}} &  \rotatebox{90}{\textbf{\textit{Hpye}}} \\ 
\midrule

\multirow{7}{*}{\rotatebox{90}{\textbf{BRC-tokens}}}  & \textbf{IV.\ding{202}} & \cellcolor{gray!10} Interest period & \cmark & \cmark & \cmark & - & - &  - & - &- & \tikz\pic{sema=green/180/white}; & \tikz\pic{sema=red/90/white}; \\ 

& \textbf{IV.\ding{203}} & \cellcolor{gray!10}  Market share & \cmark & - & \cmark & - &- & - & - &- & \tikz\pic{sema=green/270/white}; & \tikz\pic{sema=red/180/white}; \\ 

& \textbf{IV.\ding{204}} & \cellcolor{gray!10}  Returns &   \cmark &  \cmark &  - & - & - & - & - &- &  \tikz\pic{sema=green/90/white}; & \tikz\pic{sema=red/180/white}; \\ 

& \textbf{IV.\ding{205}} & \cellcolor{gray!10} Volatility &   \cmark &  \cmark &  - &- & - & - & -&- &  \tikz\pic{sema=green/270/white}; & \tikz\pic{sema=red/90/white}; \\ 

& \textbf{IV.\ding{206}} & \cellcolor{gray!10}  Risk &   \cmark &  \cmark &  - & - & - &- & -&- &  \tikz\pic{sema=green/270/white}; & \tikz\pic{sema=red/90/white}; \\ 

& \textbf{IV.\ding{207}} & \cellcolor{gray!10} Correlation &  \cmark &  \cmark &  \cmark & - & - & - & - &- &  \tikz\pic{sema=green/90/white}; & \tikz\pic{sema=red/270/white}; \\  

& \textbf{IV.\ding{208}} & \cellcolor{gray!10} User activity &  - & - & - &  \cmark   &  \cmark & - & -  &-&  \tikz\pic{sema=green/180/white}; & \tikz\pic{sema=red/90/white}; \\ 

\midrule

\multirow{3}{*}{\rotatebox{90}{\textbf{ Senti. }}} &  \textbf{V.\ding{202}} & \cellcolor{gray!10}  Sentiment & -  & - & - &  \cmark & - & \cmark & -& - & \tikz\pic{sema=green/180/white}; & \tikz\pic{sema=red/180/white}; \\ 

&  \textbf{V.\ding{203}} & \cellcolor{gray!10} User diversity  & -  & - &  - &  \cmark & - & \cmark   & - &  - & 
 \tikz\pic{sema=green/270/white}; & \tikz\pic{sema=red/180/white}; \\  

&  \textbf{V.\ding{204}} & \cellcolor{gray!10} (non-)Scam & -  & - &  - & \cmark & - & \cmark & - & -  & \tikz\pic{sema=green/270/white}; & \tikz\pic{sema=red/180/white}; \\  

\midrule

\multirow{2}{*}{\rotatebox{90}{\textbf{Hist.}}}& \textbf{VI.\ding{202}} & \cellcolor{gray!10}  Tokenwave  & - &  \cmark &  \cmark & - & - & -& - & - &  \tikz\pic{sema=green/180/white}; & \tikz\pic{sema=red/180/white}; \\  

& \textbf{VI.\ding{203}} & \cellcolor{gray!10}  Macro trend  &  - &  \cmark &  \cmark & - & - & - & - & -&  \tikz\pic{sema=green/180/white}; & \tikz\pic{sema=red/180/white}; \\   

\midrule

\multirow{3}{*}{\rotatebox{90}{\textbf{Compr.}}}& \textbf{VII.\ding{202}} & \cellcolor{gray!10}  standard, NFT &  - &  - &  - & - & - & - & \cmark &  \cmark &  \tikz\pic{sema=green/180/white}; & \tikz\pic{sema=red/90/white}; \\   

& \textbf{VII.\ding{203}} & \cellcolor{gray!10} Design pros &  - &  - &  - & - & - & - & \cmark &  \cmark &  \tikz\pic{sema=green/180/green}; & \tikz\pic{sema=red/360/white}; \\   

& \textbf{VII.\ding{204}} & \cellcolor{gray!10} Design cons  &  - &  - &  - & - & - & - & \cmark & \cmark &  \tikz\pic{sema=green/360/white}; & \tikz\pic{sema=red/180/red}; \\   

\midrule

\multicolumn{1}{c}{} & & &  \multicolumn{8}{c|}{\cellcolor{gray!10}\textbf{\textit{Indicative factors}}} & \multicolumn{2}{c}{\cellcolor{gray!10}\textbf{\textit{27, 34}}}\\ 
\end{tabular}
}
\begin{tablenotes}
      \footnotesize
      \item[] \quad\quad\quad\quad\quad\quad\quad\quad\quad\quad\quad Notably, each sector (black area, \tikz\pic{sema=black/270/white};) counts 1.
     \end{tablenotes}
\end{table}

\smallskip
\noindent\textbf{Our results.} We present a series of significant findings from each investigated section, which we synthesize in \underline{Tab.\ref{tab-result}}. Additionally, we offer an assessment of the level of both \textit{hope} and \textit{hype} within the BRC-20 ecosystem. In this context, \textit{hope} signifies the potential for sustainable prosperity, whereas \textit{hype}  denotes a surge in interest driven by arbitrage, often accompanied by a risk of overvaluation. Upon comprehensive evaluations, we observe a slight predominance of the \textit{hype}  (34) aspect over the \textit{hope} (27) element.  This suggests that a more cautious sentiment towards this new concept should be taken into consideration. Meanwhile, it's important to note that the benchmark for our analysis is ERC-based markets (including BNB Chain, Avalanche, etc.), which may lead to a certain level of imbalance when comparing Bitcoin-related markets.


\smallskip
\noindent\textbf{\textcolor{red}{\dangersign}} \textbf{Limitations.}  Our investigations have certain limitations with respect to data collection. First, we acknowledge the \textit{limited scope of our token portfolio}, which may introduce bias into our results. This limitation arises from our focus on a selected group of representative tokens, potentially excluding relevant others. The rationale behind this selection is that many tokens and projects exhibit strong correlations that might not necessarily contribute significantly to the overall market trends. Additionally, some tokens possess relatively low market capitalization and therefore may have limited impact on the broader market dynamics. Second, our analysis is constrained by \textit{the short timeframe of tweet data} collection. Due to resource constraints (majorly costs and human efforts), we conducted investigations over  a randomly chosen week of recent tweets. While this data snapshot may not capture the entire range of market sentiments, it can still provide a reasonably representative picture of recent market performance. Furthermore, our assessment is partially based on \textit{subjective summaries} and informal surveys. We remind the potential for slight inaccuracies in this analysis, particularly on the market side, which is influenced by a multitude of factors.

\smallskip
\noindent\textbf{Related sources.} Rodarmor~\cite{bipord} introduced a scheme for assigning serial numbers to Bitcoin satoshis. A relatively complete introduction to ordinal theory can be found at~\cite{ordinbook}. Binance Research published several early reports~\cite{binancebrc}\cite{binance1}\cite{binance2} that delve into the development of BRC-20. Investigating the impact of Bitcoin Ordinals on transaction fees, Bertucci~\cite{bertucci2023bitcoin} concluded that ordinal inscriptions tend to incur lower fees compared to regular transactions. In parallel, Kiraz et al.~\cite{kiraz2023nft} presented an alternative approach to settling NFT trades on the Bitcoin blockchain using zero-knowledge proofs, distinct from the ordinal method. Additionally, various media outlets have offered accessible explanations of this emerging concept~\cite{news1}\cite{news2}\cite{news3}\cite{news4}. Trevor.btc et al. have provided detailed coverage of the development of Ordinals/BRC-20 and hosted ``The Ordinal Show''~\cite{pshow} podcast.
Readers keen on further exploration can conduct searches using relevant keywords such as \textit{BRC-20}, \textit{Bitcoin NFT}, and \textit{Ordinals}, along with associated techniques covering \textit{UTXO}~\cite{antonopoulos2017mastering}, \textit{Taproot}~\cite{taproof} and \textit{SegWit}~\cite{segwit} (cf. Sec.\ref{sec-construction}) and surrounding applications (Sec.\ref{subsec-application}).



\section{BRC-20 Construction}
\label{sec-construction}

\subsection{Prelinminary: Bitcoin UTXO \& Transaction Fundamentals}\label{subsec-const-utxo}

We begin by introducing the fundamental concept of the Unspent Transaction Output (UTXO) model, which serves as the underlying framework for Bitcoin transactions. In this model (Listing~\ref{list-utxo}), the outputs of one transaction become the inputs for subsequent transactions, creating a continuous chain of transactions without the need for traditional accounts.

\begin{lstlisting}[caption={UTXO transaction model\label{list-utxo}},basicstyle=\ttfamily\scriptsize]]
Tx0 (output1: 0.5btc) ---> Tx2 (input1: 0.5btc) 
Tx2 (output1: 0.3btc) ---> Tx3 (input1: 0.3btc) 
Tx1 (output1: 0.2btc) ---> Tx2 (input2: 0.2btc) 
Tx2 (output2: 0.2btc, coinbase, output3: 0.1btc, coinbase)
Tx1 (output2: 0.1btc)
\end{lstlisting}

Each transaction is composed of inputs and outputs, where inputs refer to the outputs of previous transactions. In the UTXO model, the term \textit{fee} is used to define the difference between the total input and output amounts, which is then given to the miner who includes the transaction in a block.

Security in Bitcoin transactions is upheld by locking and unlocking scripts. The locking script (or $\mathsf{scriptPubKey}$) sets the conditions that must be met to spend the output. On the other hand, the unlocking script (or $\mathsf{scriptSig}$) is provided by the spender to meet these conditions and spend the output. It's also important to remember that 1 Bitcoin (BTC) equates to $10^8$ satoshis. As miners prioritize transactions with a higher fee rate ($\mathsf{fee\_rate = fee / size}$), the block size is typically restricted to approximately 1MB.

\subsection{Bitcoin Ordinals: Tracking Every Satoshi}\label{subsec-const-ordinal}
The second key step is achieving field uniqueness in BRC-20 by leveraging Bitcoin Ordinals, which index each satoshi based on its mining order. For example, the first-ever mined satoshi in the genesis block is indexed as 0 and can be accessed at \url{https://ordinals.com/sat/0}. Ordinals provide versatility with multiple representation formats:

\begin{itemize}
    \item \textit{Integer notation}: The ordinal number itself, reflecting the order in which the satoshi was mined. For example, 2099994106992659.

    \item \textit{Decimal notation}: The block height at which the satoshi was mined, followed by the offset within the block. For example, 3891094.16797.

    \item \textit{Degree notation}: The last number is the order in which the sat was mined in the block, followed by the block height in degrees, such as 3°111094’214’’16797’’’.

    \item \textit{Percentile notation}: The position of the satoshi in Bitcoin's total supply, expressed as a percentage. For example, 99.99971949060254\%.

    \item \textit{Name}: An encoding of the ordinal number using the characters ``$\mathsf{a}$''-``$\mathsf{z}$'', such as ``$\mathsf{satoshi}$''.
\end{itemize}

The FIFO (First-In-First-Out) principle applies once a satoshi becomes part of a transaction. Suppose a transaction involves two inputs, each containing three satoshis, and an output containing four satoshis. In that case, the output will include the first four satoshis from the combined inputs. As in Listing~\ref{list-fifo}, each ``$[...]$'' represents an input or output, and each satoshi is indexed with a character from  ``$\mathsf{a}$'' through ``$\mathsf{z}$''.

Fees are handled similarly. If a transaction has two inputs, each containing two satoshis, and one output containing three satoshis, the output will comprise the first three satoshis from the combined inputs, and one satoshi will be used as a fee and assigned to a Coinbase transaction.

\begin{lstlisting}[caption={Tracking the tagged satoshi - FIFO\label{list-fifo}},basicstyle=\ttfamily\scriptsize]]
[a b c] [d e f] --> [a b c d] [e f]
[a b] [c d] --> [a b c] [d]
Coinbase tx: [SUBSIDY] [d] --> [SUBSIDY d]
\end{lstlisting}

Within Bitcoin Ordinals, another noteworthy innovation emerges in the form of \textit{rare satoshis}~\cite{raresociety}, pursuing the most significant milestones in satoshis, similar to the iconic example of \textit{Bitcoin Pizza}~\cite{pizza}. These satoshis can be distinctly identified as having been mined from specific blocks.

\begin{itemize}
    \item \textit{Common}: Any that is NOT the first satoshi of its block.
    \item \textit{\textcolor{uncommon}{Uncommon}}: The first satoshi of each block.
    \item \textit{\textcolor{rare}{Rare}}: The first of each difficulty adjustment period.
    \item \textit{\textcolor{epic}{Epic}}: The first satoshi of each halving epoch.
    \item \textit{\textcolor{legendary}{Legendary}}: The first satoshi of each cycle.
    \item \textit{\textcolor{mythic}{Mythic}}: The first satoshi of the genesis block.
\end{itemize}

\subsection{Inscriptions: Embedding Messages in Satoshis}\label{subsec-const-inscript}

The third crucial step involves incorporating personalized content into each unique satoshi. This concept is known as \textit{Inscriptions}. Inscriptions leverage the Ordinals protocol, enabling the direct embedding of content (details in \underline{Tab.\ref{tab:inscription}}) into a satoshi in the form of JSON (JavaScript Object Notation, also refer to Sec.\ref{subsec-brcfunc}). This transformation effectively turns satoshis into NFTs, making them vessels for arbitrary data.

The data is stored within the segregated witness (SegWit~\cite{antonopoulos2017mastering}) section of a transaction. SegWit is a protocol upgrade that enhances scalability by modifying how data is stored in a block. In SegWit-enabled transactions, the transaction size ($\mathsf{tx_{size}}$) is calculated as the sum of the transaction body size and a quarter of the witness size~\cite{bertucci2023bitcoin}. The storage space is around $3–4$ MB, sufficient to accommodate text, speech, audio, images, and even highly compressed video files.

Inscriptions utilize the Taproot upgrade to Bitcoin, which serves for privacy and flexibility in transactions. For this purpose, an output designated to store an inscription is assigned the Pay-To-Taproot (P2TR) script type, and the inscription is embedded in the Taproot script inside the transaction's witness. We present a Taproot script with an inscription (Listing~\ref{list-taproot}). 

\begin{lstlisting}[caption={Taproot script\label{list-taproot}},basicstyle=\ttfamily\scriptsize]] %\\a Script Type defined in witnesses
OP_FALSE
OP_IF
OP_PUSH "ord"
OP_1
OP_PUSH "text/plain;charset=utf-8"
OP_0
OP_PUSH "Hello, world!"
OP_ENDIF
\end{lstlisting}

Building upon these techniques, ordinal-based Bitcoin NFTs stand apart from Ethereum NFTs due to three distinctions (more details refer to \underline{Tab.\ref{tab:nfts}}):

\begin{itemize}
\item  \textit{Indivisibility}. Every Bitcoin inscription is anchored to satoshis, with the issuance of these satoshis bounded by the upper limit of the Bitcoin network. The quantity of inscribed NFTs is confined to approximately $21m \times 10^{8}$.

\item \textit{Numbered sequentially}. 
Each Inscription is systematically allocated a position as outlined by the Ordinal Theory. This introduces a distinct characteristic capable of conferring diverse levels of value upon distinct sequential creations, including Inscriptions minted following the block reward halving or the inaugural Inscription itself.

\item \textit{Scale limitation}. 
The Bitcoin block can accommodate a maximum of 4MB of data after the SegWit and Taproot upgrades. Considering that approximately 144 Bitcoin blocks can be mined daily, a total of about 210GB of space is available annually for Inscription minting (a single Inscription requires 4MB of space). In contrast,  NFTs based on smart contracts lack such limitations, theoretically allowing for unlimited minting.
\end{itemize}

\subsection{Extension: ORC-20 and Surroundings}
\label{subsec-application}

\begin{wraptable}{r}{4.5cm}
\vspace{-0.1in}
\setlength{\abovecaptionskip}{-0.05in}
\caption{Inscription contents}\label{tab:inscription}
\renewcommand\arraystretch{1.1}
\begin{center}
   \begin{threeparttable}
\resizebox{\linewidth}{!}{
\begin{tabular}{l|r} 
\toprule
        \textbf{Standard} & \textbf{Special inscriptions}  \\
\midrule
        text / files &  \cellcolor{gray!10}from erc721 collection \\
        sats domains & \cellcolor{gray!10}sats domain hosting \\
        unisat domains & \cellcolor{gray!10} from solana  \\
        brc-20 mint &  \cellcolor{gray!10}orc-20 mint \\
        brc-20 deploy &   \cellcolor{gray!10}orc-20 deploy \\
        brc-20 transfer & \cellcolor{gray!10}loot for taproot \\
\bottomrule
\end{tabular}
}
     \end{threeparttable}
     \end{center}
     \vspace{-0.11in}
\end{wraptable}

\noindent\textbf{OCR-20.} ORC-20 \cite{orc20}, created by OrcDAO, is an open standard designed to enhance the capabilities of ordered tokens on the Bitcoin network. It ensures seamless backward compatibility with BRC-20. Unlike BRC-20, which necessitates a \textit{one-time transfer inscription} in each transaction, ORC-20 allows for the reusability of the \text{mint} and \textit{send} ordinal inscriptions within a transaction.

\smallskip
\noindent\textbf{Surroundings.} We also investigate a series of supporting applications that are relevant to BRC-20 (\underline{Tab.\ref{tab:tool}}).

\begin{table}[!hbt]
\caption{BRC-20/Ordinals Surrounding Tools}\label{tab:tool}
\vspace{-0.15in}
\begin{center}
\resizebox{1\linewidth}{!}{
\begin{tabular}{c|r} 
\toprule
\cellcolor{gray!10}\textit{\text{Wallet}}     &  UniSat, Ordinals, Xverse, Hiro, Sparrow, Bitcoin.com\\ 
\cellcolor{gray!10}\textit{\text{Mint}}     & Gamma.io, Ordinals Bot \\
\cellcolor{gray!10}\textit{\text{Tool}}     &  Rarity Garden, OrdSpace, BestinSlot, Mempool \\
\cellcolor{gray!10}\textit{\text{Market}}     &   Magic Eden, OpenOrder, Ordswap, Twetch, Openordex\\ 
\cellcolor{gray!10}\textit{\text{NFT}} &TwelveFold, Bitcoin Rocks, Taproot Wizards, Pixel pepes  \\ 

 \bottomrule
\end{tabular}
}\end{center}
\end{table}



\section{BRC-20 on Bitcoin Networks}
\label{sec-protocol}

\subsection{Implementing BRC-20}
\label{subsec-brcfunc}

The design of implementation is to address the incompatibility between the stateless UTXO-based models of Ordinals and the stateful account-based approach of BRC-20. At the heart of this reconciliation is the use of inscriptions to record state transitions, transforming these immutable markers into auditable proofs. This method hinges on the construction and maintenance of an \textit{off-chain state indexer}, which records the balance of each account. Inscriptions on the Bitcoin network then serve as triggers to update these off-chain states. In essence, BRC-20 has enabled three primary functions. 

\smallskip
\noindent\textcolor{teal}{\ding{49} \textit{Deploy a new token.}} The operation initiates the creation of a new BRC-20 token (\textcolor{teal}{$\mathsf{Deploy}$}, Listing~\ref{list-deploy}). It begins on-chain with the inscription of a satoshi to represent the deployment. This inscription contains several crucial details such as the protocol name (\textit{brc-20}), operation (\textit{deploy}), token's name (\textit{tick}), the total amount of tokens to be issued (\textit{max}), and the maximum amount of tokens to be minted in each minting round (\textit{lim}). After this inscription is added to the Bitcoin network, an off-chain process verifies whether a state already exists for the given token name. If not, a new state is created, with the balance of each account initialized to zero or a pre-defined value and the token's properties (those defined in Inscriptions) added to the state. The on-chain inscription structure and the off-chain update are listed below.

\begin{lstlisting}[caption={ \textcolor{teal}{$\mathsf{Deploy}$}\label{list-deploy} },basicstyle=\ttfamily\scriptsize]]
# Onchain Inscription
"p" : "brc-20", # protocol name
"op": "deploy", # operation
"tick": "ordi", # token name
"max": "2100000", # total amount of token to be issued
"lim": "1000" # maximum amount of token minted each round

# Off-chain update
if state[tick] NOT exists:
    state[tick]={<inscription info>, 
                 "balances": {addr: balance_val}}
\end{lstlisting}


\noindent\textcolor{teal}{\ding{49} \textit{Mint new tokens.}}  This operation is to issue new tokens (\textcolor{teal}{$\mathsf{Mint}$}, Listing~\ref{list-mint}). Similar to \textcolor{teal}{$\mathsf{Deploy}$}, this operation begins on-chain with the inscription of a satoshi to represent the mint operation. The inscription contains the protocol name (\textit{brc-20}), the operation (\textit{mint}), the token name, and the number of tokens being minted (\textit{amt}). Once the inscribed satoshi is transferred to the minter, an off-chain process takes over. This process verifies whether a state already exists for the token, checks that the number of tokens minted does not exceed the limit defined in the \textcolor{teal}{$\mathsf{Deploy}$} operation, and ensures that the total number of minted tokens does not surpass the maximum amount. If all checks pass, the minter's balance in the off-chain state is increased by the number of newly minted tokens. 
\begin{lstlisting}[caption={\textcolor{teal}{$\mathsf{Mint}$}\label{list-mint}},basicstyle=\ttfamily\scriptsize]]
# Onchain Inscription
"p" : "brc-20", # protocol name
"op": "mint", # operation
"tick": "ordi", # token name
"amt": "1000" # the amount of token being minted

# Off-chain update
if state[tick] NOT exists OR 
                 "amt" > "lim" OR sum("amt") > "max":
    raise errors
else 
    account_state[tick]["balance"][minter] += amt
\end{lstlisting}

\noindent\textcolor{teal}{\ding{49} \textit{Transfer tokens for trade.}}  The operation facilitates the movement and exchange of tokens between accounts (\textcolor{teal}{ $\mathsf{Transfer}$}, Listing~\ref{list-transfer}). This operation is slightly more complicated due to the need for two on-chain transactions. First, a satoshi is inscribed to represent the transfer operation. The inscription includes the protocol name (\textit{brc-20}), the operation (\textit{transfer}), the token name, and the amount being transferred. This inscribed satoshi is then transferred to the sender, who needs to immediately create a second transaction to forward the inscribed satoshi to the recipient. On the off-chain side, the sender's balance is verified to ensure that it covers the transfer amount. If it does, the sender's balance is decreased, and the recipient's balance is increased correspondingly.

\begin{lstlisting}[caption={ \textcolor{teal}{$\mathsf{Transfer}$}\label{list-transfer} } ,basicstyle=\ttfamily\scriptsize]]
# Onchain Inscription
"p" : "brc-20", # protocol name
"op": "transfer", # operation
"tick": "ordi", # token name
"amt": "100" # the amount of token being transferred

# Off-chain update
if state[tick] NOT exists:
    raise errors
if state[tick]["balance"][sender] >= amt:
    account_state[tick]["balance"][sender] -= amt
    account_state[tick]["balance"][receiver] += amt
\end{lstlisting}

\subsection{Operating BRC-20 (NFT) on Bitcoin}

\noindent\textbf{The PSBT standard.} PSBT, short for partially signed Bitcoin transactions, is a Bitcoin standard (BIP-174~\cite{orc20}) that enhances the portability of unsigned transactions and enables multiple parties to easily sign the same transaction.
A PSBT is created with a set of UTXOs to spend and a set of outputs to receive. Then, the information of each UTXO necessary to create a signature will be added. Once the PSBT is prepared, it can be copied to a program capable of signing it. For multi-signature wallets, this signing step can be repeated using different programs on separate PSBT copies. Multiple PSBTs, each containing one or more necessary signatures, will later be combined into a single PSBT. Finally, the fully signed PSBT can be broadcast via networks.

\smallskip
\noindent\textbf{Transaction workflow.} Building upon this standard, we present a complete cycle for trading a BRC-20  transaction.

\smallskip
\noindent\textcolor{teal}{\ding{49} \textit{Seller's Operation.}}  
    A seller uses a transaction to inscribe a satoshi, indicating a transfer operation of a certain amount of BRC-20 tokens (e.g., \textit{1000 ordi}). The inscribed satoshi manifests the seller's intent to sell the stated amount of tokens and carries detailed information, including the protocol name (\textit{brc-20}), the operation (\textit{transfer}), the token name (\textit{ordi}), and the transfer amount (e.g., \textit{1000}).

\noindent\textcolor{teal}{\ding{49} \textit{Creation of PSBT.}} 
    Next, the seller incorporates the inscribed satoshi as an input in PSBT. To set the starting bid, the seller designates an output in the PSBT for \textit{the seller transfers 0.2 BTC to their own address.} This action signifies the seller's intention to exchange \textit{1000 ordi} tokens for \textit{0.2 BTC}.

\noindent\textcolor{teal}{\ding{49} \textit{Publishing the PSBT.}} 
    Then, the seller publishes the PSBT to a marketplace, allowing potential buyers to review the transaction details and decide whether they wish to proceed.

\noindent\textcolor{teal}{\ding{49} \textit{Buyer's Operation.}} 
    If a buyer finds the \textit{1000 ordi} package appealing, they can select and finalize this PSBT. It indicates the buyer is willing to complete the exchange by providing the required funds (\textit{0.2 BTC} in this case) and, in return, receiving the inscribed satoshi from the seller.

\noindent\textcolor{teal}{\ding{49} \textit{Finalizing the PSBT.}} 
    Upon completing the PSBT, the buyer broadcasts it to the Bitcoin network. This entails sending the transaction data to the network, where it will be included in a future block and ultimately confirmed. Once included in a block, the transaction becomes visible to all network participants and becomes irreversible.
    
\noindent\textcolor{teal}{\ding{49} \textit{Off-chain State Updates.}} 
     After the on-chain finalization of the PSBT, the off-chain states need to be updated to reflect the new balances of the buyer and the seller.  The buyer's \textit{ordi} token balance increases by \textit{1000}, while the seller's \textit{ordi} token balance decreases by the same amount. Simultaneously, the seller's Bitcoin balance increases by \textit{0.2 BTC}, while the buyer's Bitcoin balance decreases accordingly. 

\smallskip
It is worth noting that the protocol necessitates two on-chain transactions to finalize the \textcolor{teal}{$\mathsf{Transfer}$} operation, ensuring a secure settlement for the trade between sellers and buyers.

\section{Token Investigations over Months}
\label{sec-investi-token}

\noindent\textbf{Investigation overview.} 
We have specifically selected representative projects, including the foremost three BRC-20 projects (ORDI, MOON, OSHI), each boasting a market capitalization\footnote{Top BRC-20 coin explorer: \url{https://www.coingecko.com/en/categories/brc-20} [Aug 2023].} surpassing US\$10 million. Additionally, we include the top five ERC-20 projects (MATIC, SHIB, WBTC, DAI, LINK) each with a market capitalization\footnote{Top ERC-20 coin explorer: \url{https://coincodex.com/cryptocurrencies/sector/ethereum-erc20/} [Aug 2023].} exceeding US\$4 billion. Our data spans a period of four months, commencing from April (prior to the BRC craze) and extending through August (the present date of this study's composition).

\subsection{Price and Marketcaps Trends}

\begin{figure*}[!hbt]
    \centering
    \subfigure[\textbf{(Token-)Price trends} on prevalent BRC-20 and ERC-20 projects]{\label{fig:price-trend}
    \includegraphics[width=0.93\linewidth]{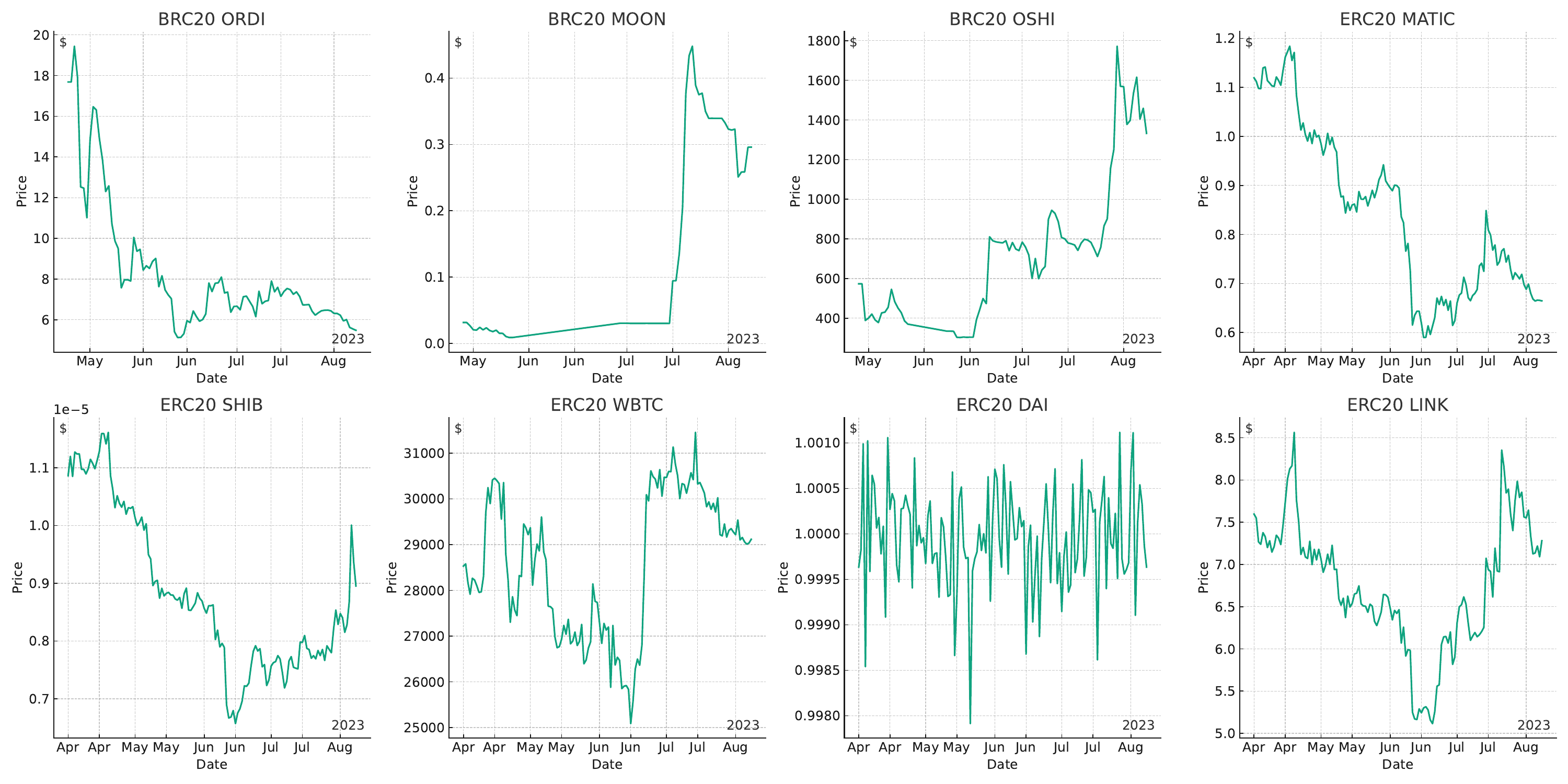}
    }
    \vspace{-0.1in}
    \subfigure[\textbf{Marketcap trends} on prevalent BRC-20 and ERC-20 projects]{\label{fig:market-cap-trend}
    \includegraphics[width=0.93\linewidth]{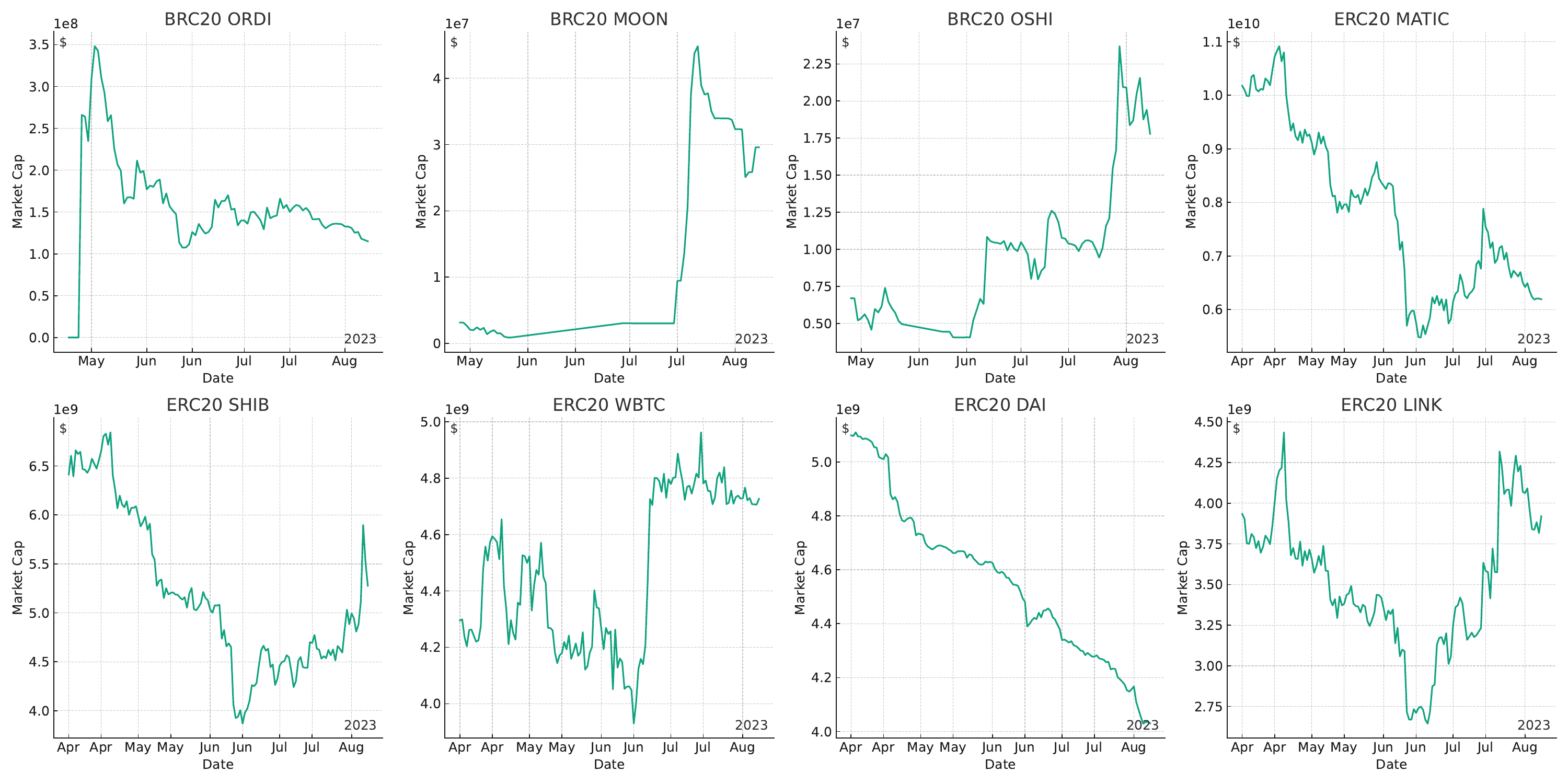}
    }
    \caption{Comparison on trends}
    \label{fig:trend}
     \vspace{-0.15in}
\end{figure*}

As \textbf{price trends} unfold (cf. Fig.\ref{fig:price-trend}), BRC-20 related tokens, represented by ORDI and MOON, exhibit sharp price increases shortly after their launch. This rapid appreciation in price is indicative of a surge in demand, likely driven by heightened market interest in these new offerings. However, such rapid increases in price can also signal overvaluation, particularly if they are not backed by strong fundamentals.

In contrast, ERC-20 related tokens, with the exception of SHIB, tend to show more stable price trends. It suggests that these coins' prices are less likely to be influenced by short-term market sentiment and more likely to reflect their intrinsic value. In particular, stablecoins like DAI can serve as a reliable store of value in the often volatile crypto markets

Examining \textbf{marketcap trends}  (see Fig.\ref{fig:market-cap-trend}), we observe substantial expansion for BRC-20 coins subsequent to their introduction. This growth is not solely attributed to price escalation; rather, it signifies increased coin circulation, implying a burgeoning user community and broader adoption of these coins. However, akin to the price dynamics, rapid market capitalization growth bears a dual nature: it can signal a coin's promise, yet it might also signify hype if the acceleration is overly swift and lacks sustainability.

\begin{figure*}[!hbt]
    \centering
    \subfigure[Average returns]{\label{fig:avg-returns}
    \includegraphics[width=0.3\linewidth]{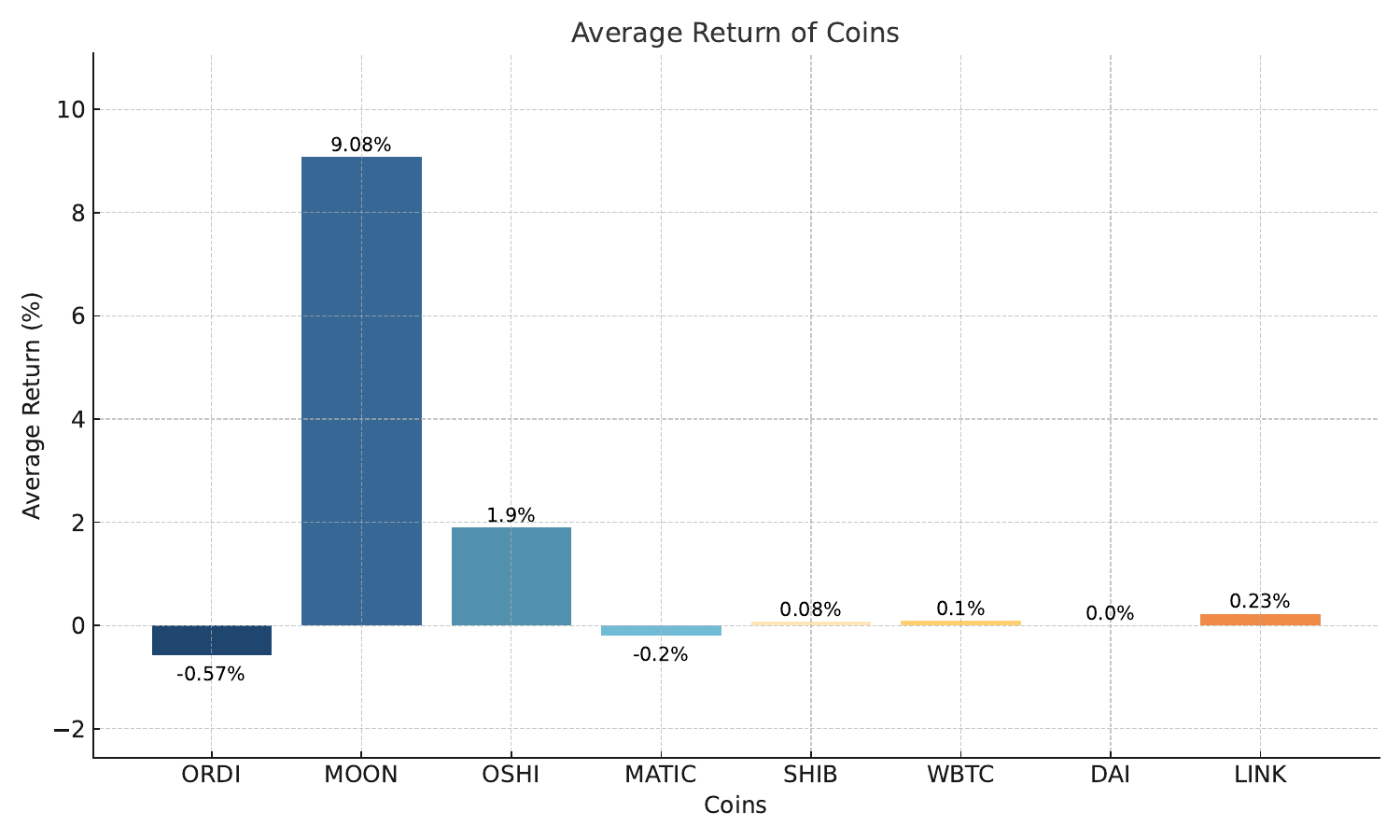}
  }
    \subfigure[Volatility]{\label{fig:volatility-analysis}
    \includegraphics[width=0.3\linewidth]{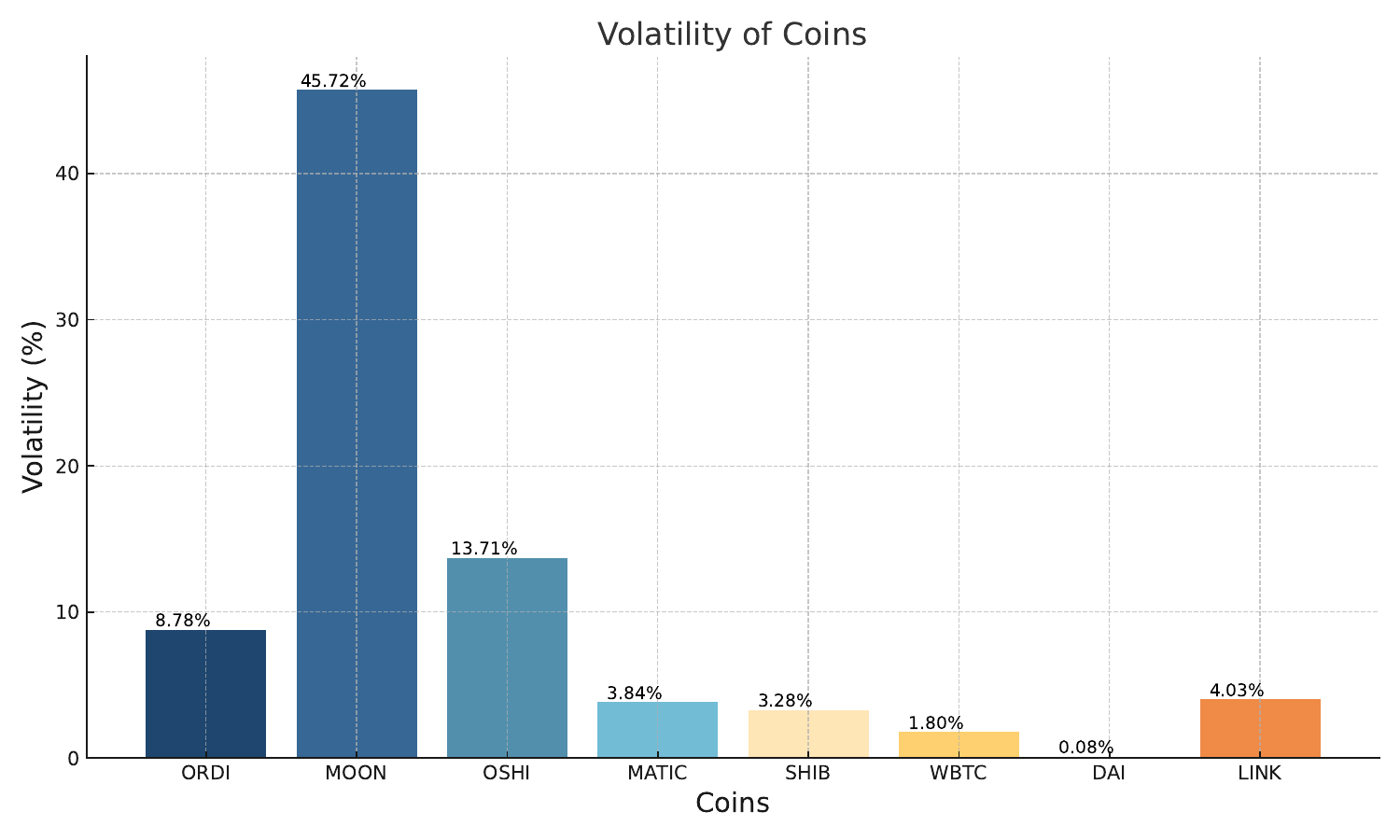}
    }
    \subfigure[Performance (Sharpe Ratio)]{\label{fig:performance-analysis}
    \includegraphics[width=0.34\linewidth]{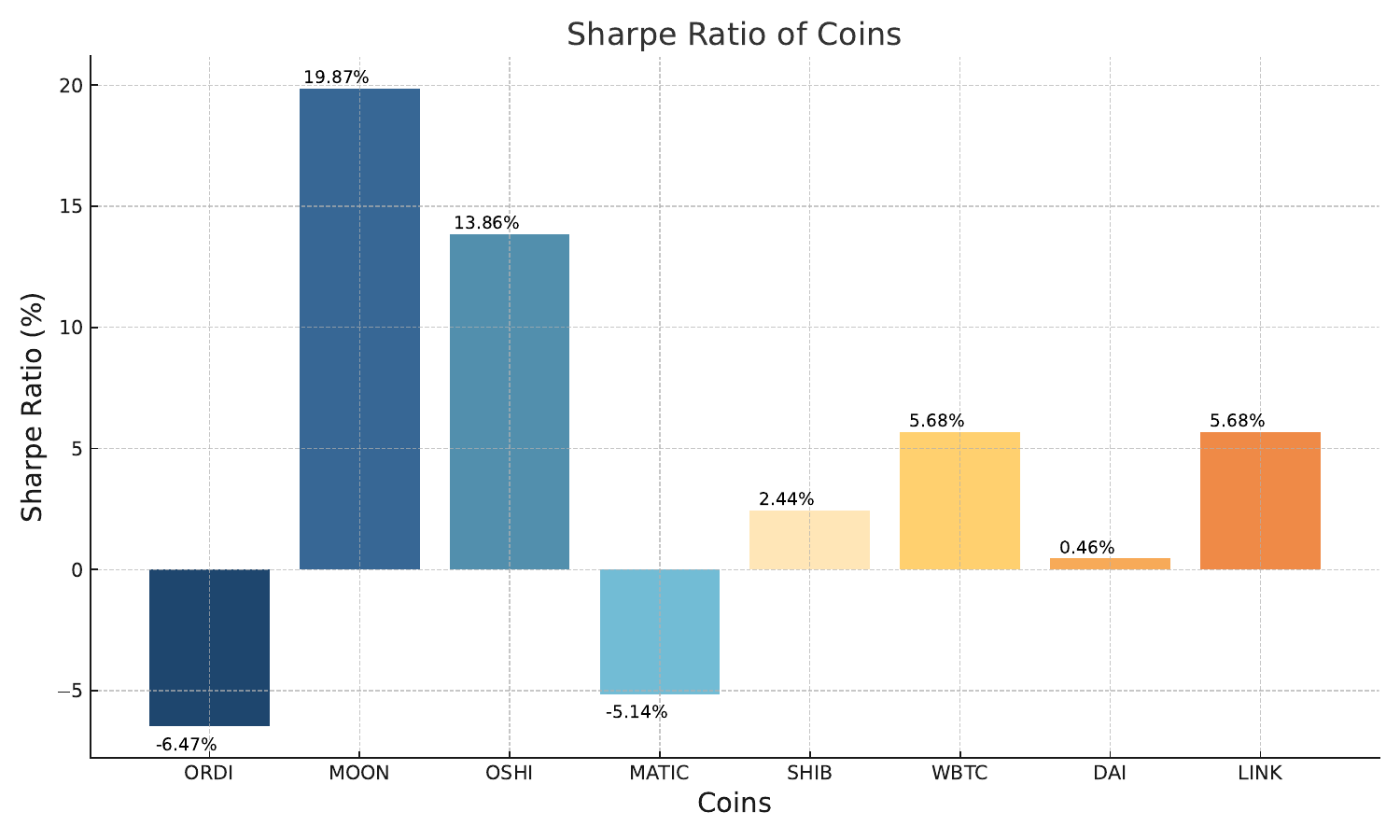}
    }
    \caption{Evaluations on prevalent BRC-20 and ERC-20 projects}
    \label{fig:evaluation}
    \vspace{-0.15in}
\end{figure*}

\smallskip
\noindent \colorbox{pink!50}{\textcolor{black}{\textbf{\text{Finding-IV.\ding{202}:}}}} \textit{\textbf{Users are rapidly entering the BRC market within a span of one month, but they may face the fact of losing their enthusiasm shortly in the subsequent months.}}

\noindent \colorbox{pink!50}{\textcolor{black}{\textbf{\text{Finding-IV.\ding{203}:}}}} \textit{\textbf{Compared to ERC-like tokens, BRC-based tokens constitute a small portion of the overall market size.}}

\subsection{Average Return}
The indicator \textbf{average return} represents the percentage change in price, serving as an indicator of profitability: a higher average return indicates greater gains for an investor who bought the coin at the beginning of the period and subsequently sold it. The chart illustrated in Fig.\ref{fig:avg-returns} visually displays the mean returns of the three BRC-20 tokens (\textit{\textcolor{blue}{blue}} bars) and the five ERC-20 tokens (\textit{\textcolor{red}{red}} bars). Evidently, the BRC-20 tokens, notably ORDI and MOON, demonstrate markedly higher average returns when compared to most ERC-20 tokens (possibly due to experiencing a high return rate during their initial launch rather than their current stable period). This suggests that over the observed duration, BRC-20 tokens may have presented an enhanced potential for profitability. It's worth noting that SHIB boasts a high return rate, aligning with the characteristics of memecoins like Dogecoin.

\smallskip
\noindent \colorbox{pink!50}{\textcolor{black}{\textbf{\text{Finding-IV.\ding{204}:}}}} \textit{\textbf{Certain BRC-20 tokens have demonstrated a remarkable return rate, often exceeding tenfold compared to equivalent tokens within the same period.}}

\subsection{Volatility Analysis}

The concept \textbf{volatility}, typically quantified as the standard deviation of returns, embodies a measure of risk: heightened volatility signifies greater price variability and consequently elevated risk. As depicted in Fig.\ref{fig:volatility-analysis}, we discern that, except for ORDI, the BRC-20 coins exhibit higher volatilities in comparison to the majority of ERC-20 coins. This observation implies that throughout the assessed period, BRC-20 coins might have entailed increased risk. This observation aligns with the earlier insight that BRC-20 coins also yielded superior returns, reinforcing the tenet that elevated returns are often accompanied by elevated risk. Conversely, with the exception of SHIB, the remaining ERC-20 tokens manifest greater stability, characterized by a narrower range of price fluctuations. We postulate that SHIB's substantial and abrupt fluctuations may stem from its memecoin attributes, rendering it particularly sensitive to market dynamics, such as significant movements instigated by prominent market participants.

\smallskip
\noindent \colorbox{pink!50}{\textcolor{black}{\textbf{\text{Finding-IV.\ding{205}:}}}} \textit{\textbf{BRC-20 tokens showcase elevated volatilities and associated risks, aligning with their substantial returns.}}

\subsection{Performance Analysis}

In our evaluation, we examine their \textbf{performance} using the Sharpe ratio\footnote{Calculated as $\mathsf{Sharpe\, Ratio} = \frac{\mathsf{Average\, Return\, of\, Investment} - \mathsf{RiskFree\, Rate}}{\mathsf{Standard\, Deviation\, of\, Investment}}$} \cite{sharpe1998sharpe}, a risk-adjusted return metric, to assess the efficacy of BRC-20 and ERC-20 tokens. The outcomes presented in Fig.\ref{fig:performance-analysis} reveal that, within the chosen tokens, both BRC-20 and ERC-20 tokens exhibit a diverse spectrum of Sharpe ratios, signaling varying levels of risk and return within these two token categories.
It shows a diverse range of Sharpe Ratios, with DAI displaying a significantly negative value, while others like SHIB and WBTC exhibit modest positive ratios. A negative Sharpe Ratio might be indicative of a high-risk, low-reward scenario, often associated with market hype and speculative trading. On the other hand, a positive Sharpe Ratio could signal a more balanced risk-reward profile, hinting at the genuine potential or ``hope'' in the investment. The presence of these dynamics in BRC-20 markets may suggest a complex landscape, where both hope and hype coexist.



\smallskip
\noindent \colorbox{pink!50}{\textcolor{black}{\textbf{\text{Finding-IV.\ding{206}:}}}} \textit{\textbf{BRC-20 tokens demonstrate heightened return rates alongside increased risks, with both the absolute values surpassing those observed in ERC-like tokens.}}

\subsection{Correlation Analysis}

\begin{figure}[!hbt]
    \centering
    \includegraphics[width=0.9\linewidth]{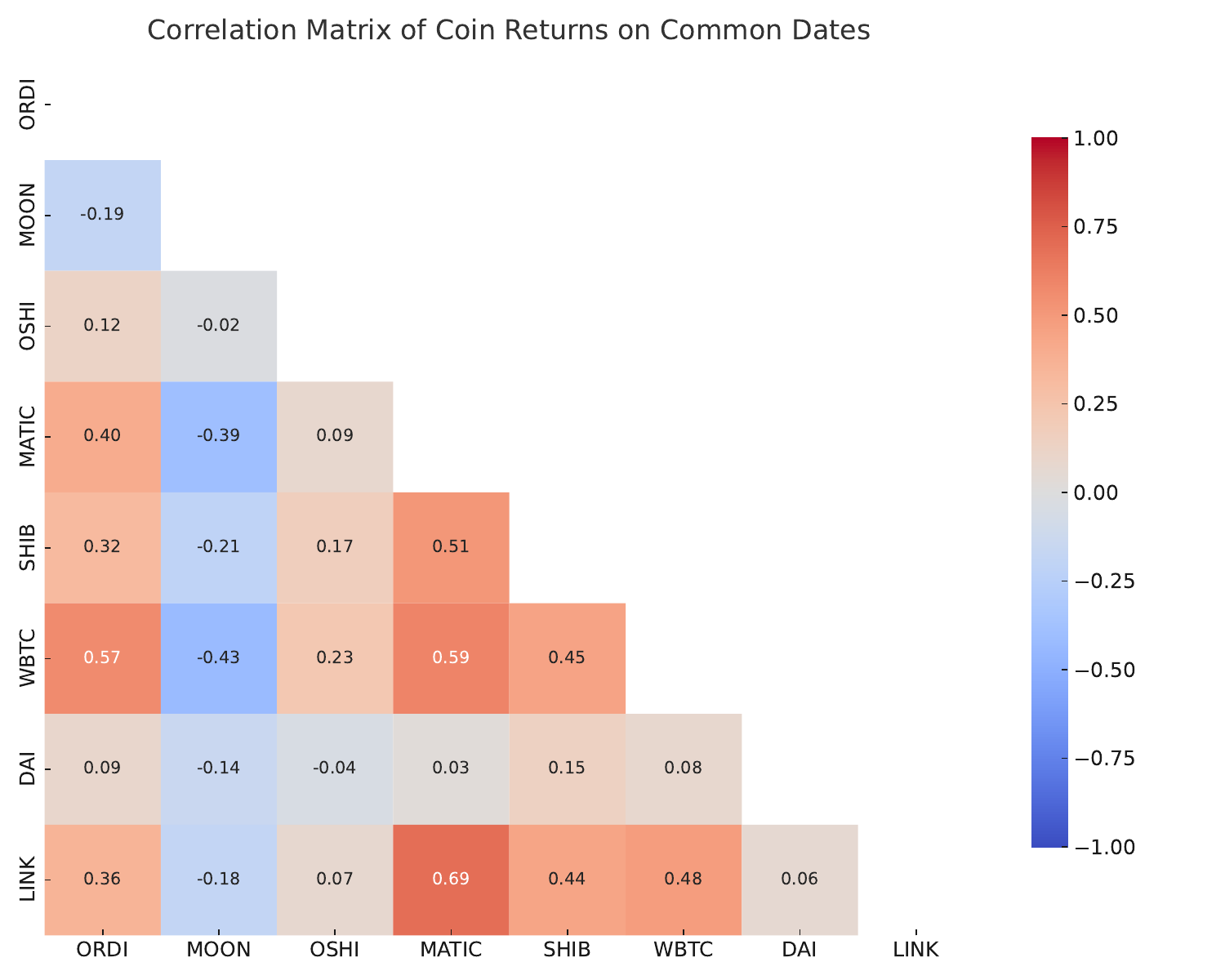}
    \caption{Correlation}
    \label{fig:correlation-analysis}
\end{figure}

The \textbf{correlation matrix} analyzing the daily returns yields insights into the relationships among chosen assets (Fig.\ref{fig:correlation-analysis}). Among the BRC-20 tokens (ORDI, MOON, and OSHI), their correlation coefficients with each other are notably elevated, indicating a robust positive linkage in their price movements. This suggests that BRC-20 tokens, as a collective, tend to exhibit synchronous shifts, possibly due to shared market perception, common underlying methodologies (rooted in ordinals), or interdependencies within the ecosystem (such as shared developers/buyers). The pronounced correlations within BRC-20 group highlight their lack of independence in a portfolio context, a crucial factor to consider in devising strategies.

Among the ERC-20 tokens (MATIC, SHIB, WBTC, DAI, and LINK), the correlation coefficients also generally exhibit positivity, albeit with less intensity compared to the BRC-20 tokens. This disparity could stem from the more established and diverse landscape of the ERC-20 token market, encompassing a wider spectrum of blockchain applications.

A comparison between these two categories unveils discernible variations in correlation coefficients. While some movements overlap, distinctive traits remain. For instance, BRC-20's ORDI demonstrates a strong positive correlation with ERC-20's LINK and WBTC, indicating a similar response to market conditions. In contrast, BRC-20's MOON exhibits a lower correlation with these ERC-20 tokens, implying distinct market dynamics at play.

\smallskip
\noindent \colorbox{pink!50}{\textcolor{black}{\textbf{\text{Finding-IV.\ding{207}:}}}} \textit{\textbf{BRC-20 tokens exhibit closely positive correlations among themselves, stronger than those in ERC-like tokens. The correlations between BRC-20 and ERC-20 tokens, however, display relatively weak connections.}}

\subsection{Usage Trend}
\begin{figure}[!hbt]
    \centering
    \includegraphics[width=1\linewidth]{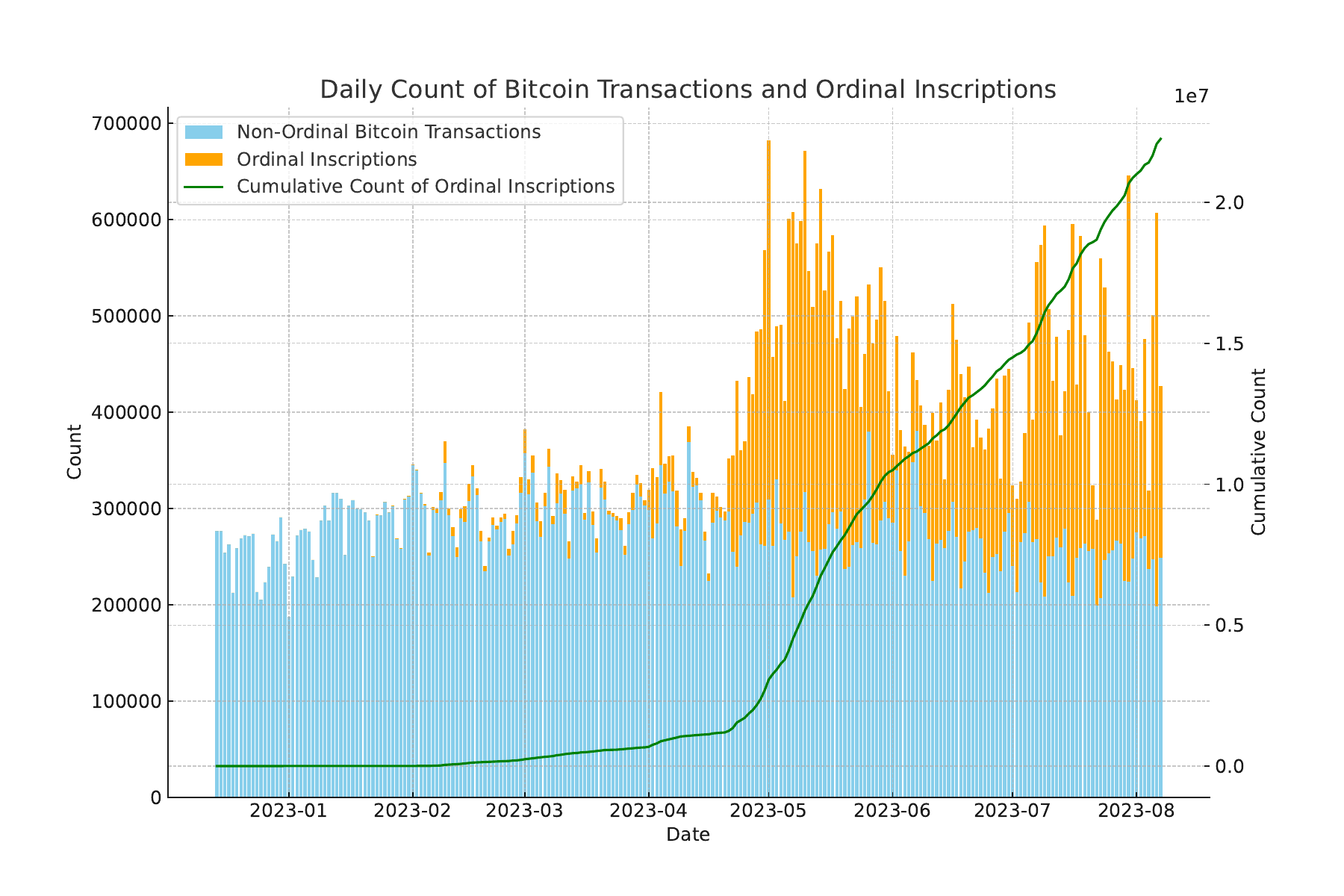}
    \caption{Daily count of BRC-20 upon Bitcoin transactions}
    \label{fig:daily-count}
\end{figure}

We proceed to compare \textbf{daily Bitcoin transactions} with \textbf{Ordinal inscriptions} (BRC-20 tokens) as depicted in Fig.\ref{fig:daily-count}. The findings reveal a steady growth in the volume of Ordinal inscriptions (\textcolor{orange}{orange} segments in bars). The cumulative count of Ordinal inscriptions (\textcolor{green}{green} line) exhibits a clear upward trajectory, indicating a progressive surge in the utilization and adoption of BRC-20 tokens over time.

However, the growth of Ordinal inscriptions should not be viewed in isolation. Non-ordinal Bitcoin transactions (\textcolor{blue}{blue} segments in bars) still form a significant portion of daily transactions. This suggests that while BRC-20 tokens are gaining traction, traditional Bitcoin transactions remain prevalent.

\smallskip
\noindent \colorbox{pink!50}{\textcolor{black}{\textbf{\text{Finding-IV.\ding{208}:}}}} \textit{\textbf{Bitcoin inscriptions have witnessed a consistent growth, yet they still represent a minor fraction of daily transactions within the overall network activity.}}

\section{Sampled Sentiment Investigations}
\label{sec-investi-sentiment}

\noindent\textbf{Investigation overview.} 
Our experiments involve gathering public tweet data from a randomly selected week (August 5th to August 9th, 2023) to delve into the prevailing perceptions and attitudes toward BRC-20. The data gathered in our experiments amounts to approximately 2 megabytes and spans interactions with around 4,112 tweet users that mentioned the hashtag \textcolor{teal}{\#brc20} or similar. We opted for this particular week as it closely aligns with the timeframe of paper composition.

\subsection{Sentiment Reactions}

We also analyze the \textbf{user sentiment} and the public perception of BRC-20 and Ordinals.

\begin{figure}[!hbt]
    \centering
    \includegraphics[width=1.1\linewidth]{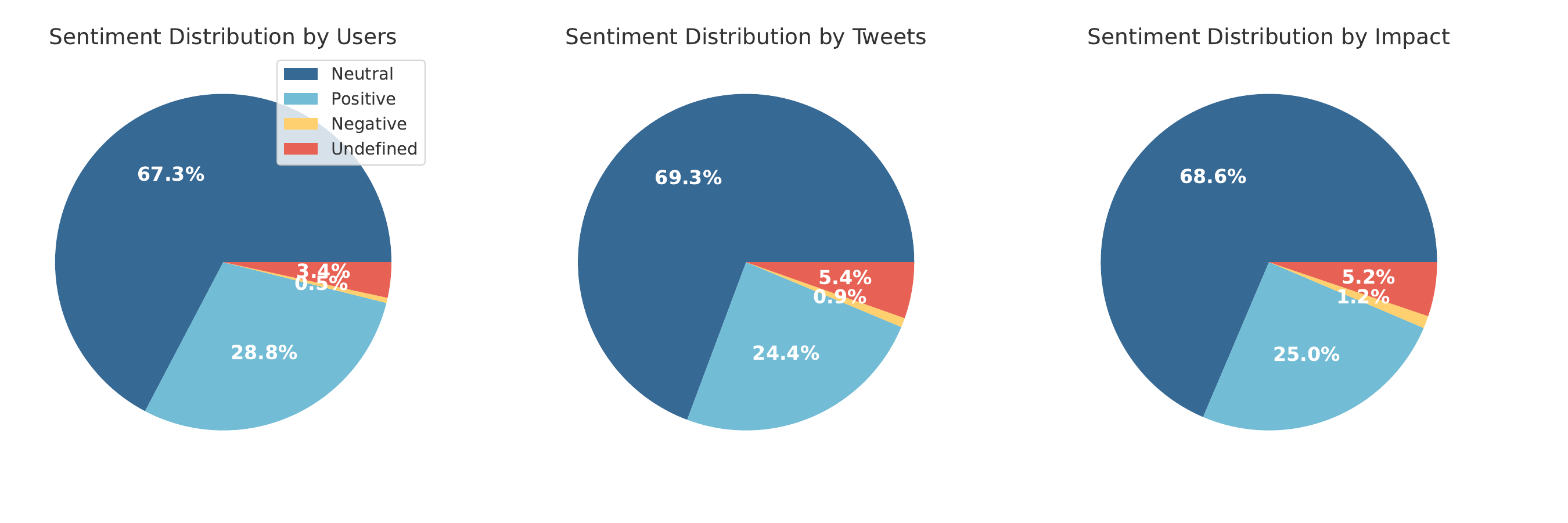}
    \caption{Sentiment distribution}
    \label{fig:senti-distri}
\end{figure}

Fig.\ref{fig:senti-distri} reveals a largely neutral sentiment across all metrics - users, tweets, and potential impact - with positive sentiment following closely behind. This distribution could be indicative of a cautiously optimistic stance toward these tokens. However, negative sentiment is minimal, comprising less than 1\% in all cases. The minimal presence of undefined sentiment suggests that most discussions are clear.

\begin{figure}[!hbt]
    \centering
    \includegraphics[width=1\linewidth]{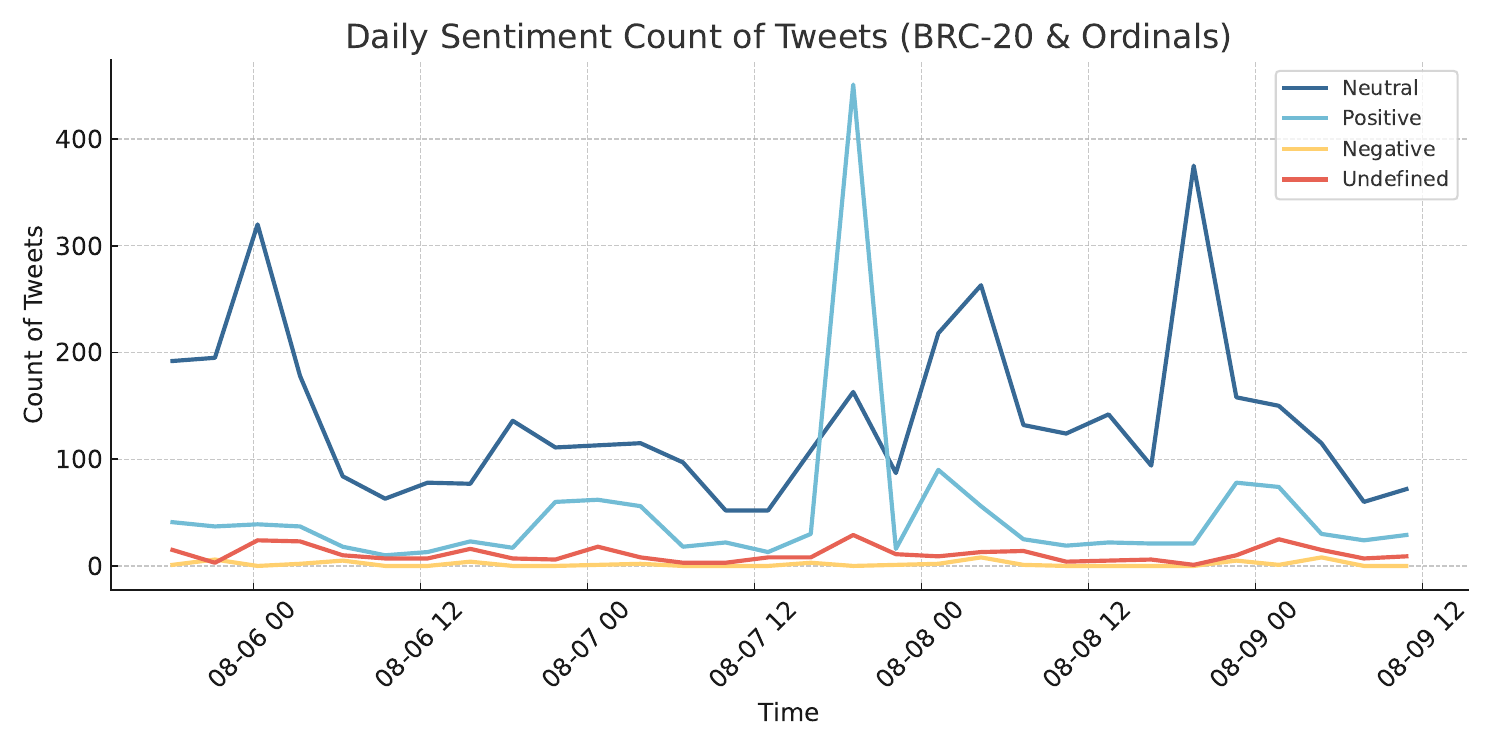}
    \caption{Sentiment actions count by Tweets}
    \label{fig:senti-tweets}
\end{figure}

Fig.\ref{fig:senti-tweets} (time series) illustrates the daily sentiment counts, showing that neutral sentiment is consistently the most prevalent, followed by positive sentiment. Negative sentiment remains relatively low throughout the investigated period. A noticeable spike in undefined sentiment around August 7 might suggest a moment of uncertainty or controversy in the discourse, but it was short-lived.

The sentiment analysis suggests that the BRC-20 and Ordinals are currently viewed more with hope than hype. The dominance of neutral and positive sentiments, coupled with the minimal negative sentiment, indicates a generally optimistic perception. Nonetheless, since our investigation timeframe is relatively brief and sentiment tends to oscillate with market dynamics, maintaining continuous monitoring would be prudent to observe any shifts in public opinion.


\smallskip
\noindent \colorbox{pink!50}{\textcolor{black}{\textbf{\text{Finding-V.\ding{202}:}}}} \textit{\textbf{Users who are inclined to express opinions have a non-negative attitude towards BRC-related concepts.}}

\subsection{Tweets with Relevant Hashtags}

\begin{figure*}[!hbt]
    \centering
    \subfigure[Tweets per contributors]{
    \begin{minipage}[t]{0.3\textwidth}
    \centering
    \includegraphics[width=2.2in]{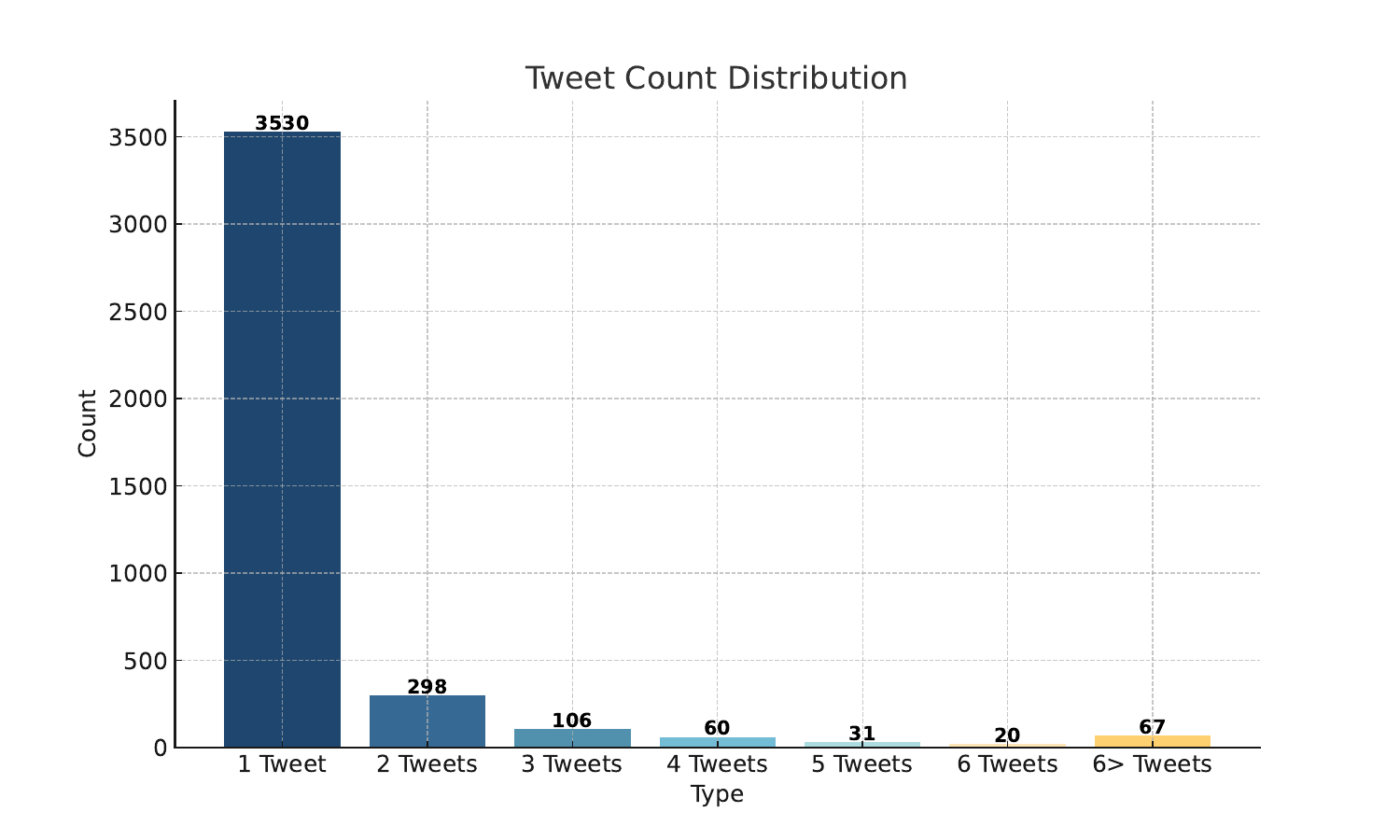}
    \end{minipage}
    \label{fig:tweets_per_contributor}
    }
    \subfigure[Source distribution]{
    \begin{minipage}[t]{0.3\textwidth}
    \centering
    \includegraphics[width=2.2in]{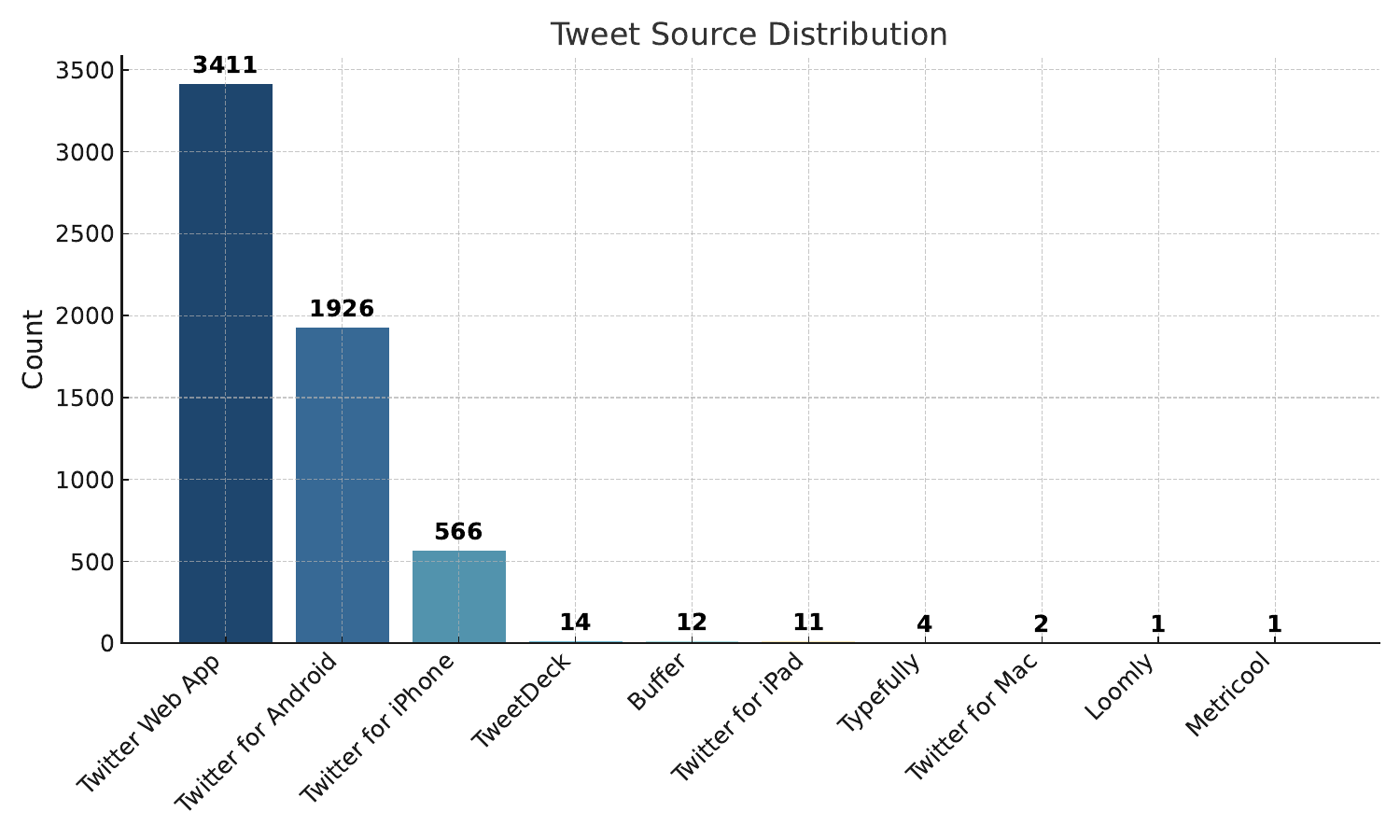}
    \end{minipage}
    \label{fig:source_distribution}
    }
    \subfigure[Age of Twitter accounts]{
    \begin{minipage}[t]{0.3\textwidth}
    \centering
    \includegraphics[width=2.2in]{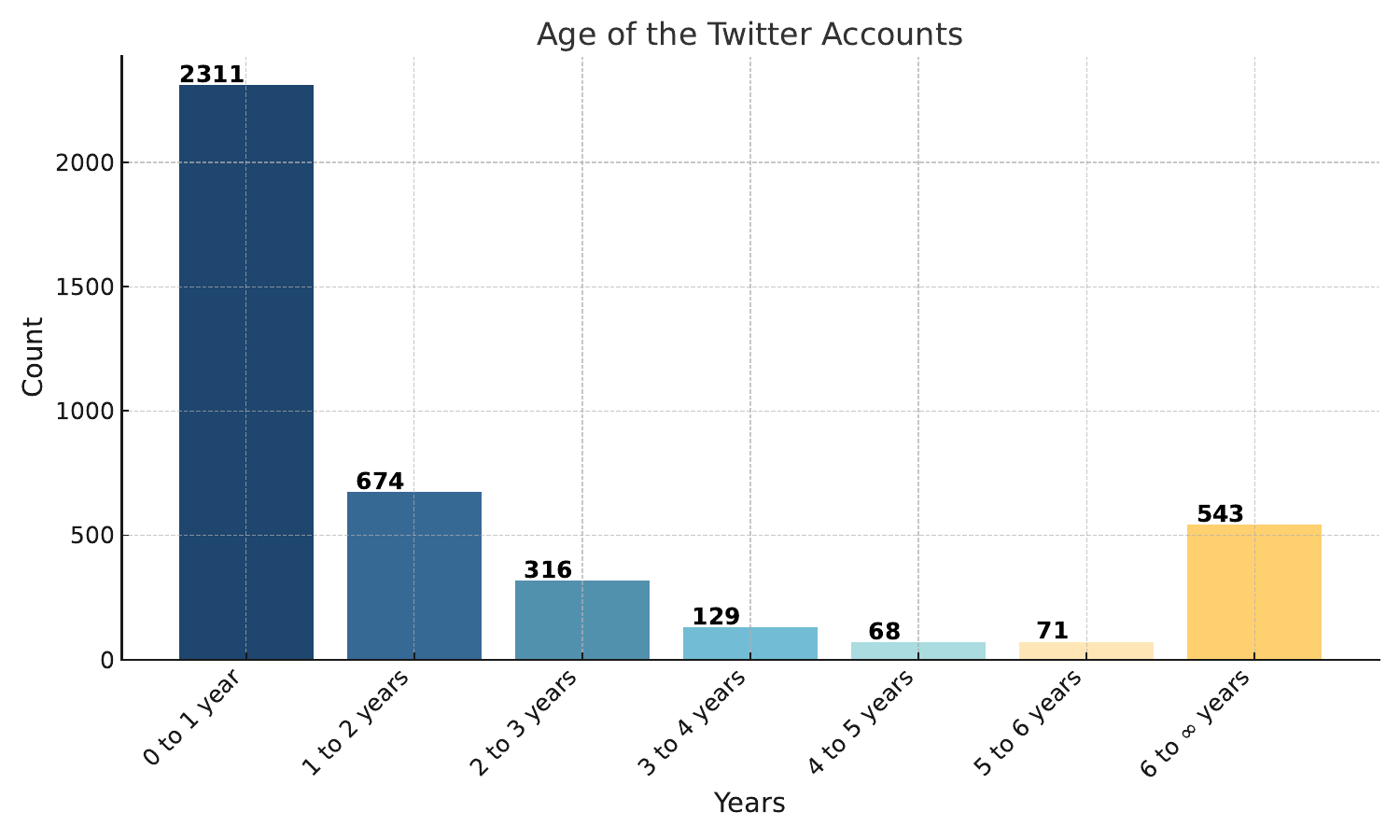}
    \end{minipage}
    \label{fig:age_of_accounts}
    }
    
    \subfigure[Length of tweets]{
    \begin{minipage}[t]{0.3\textwidth}
    \centering
    \includegraphics[width=2.2in]{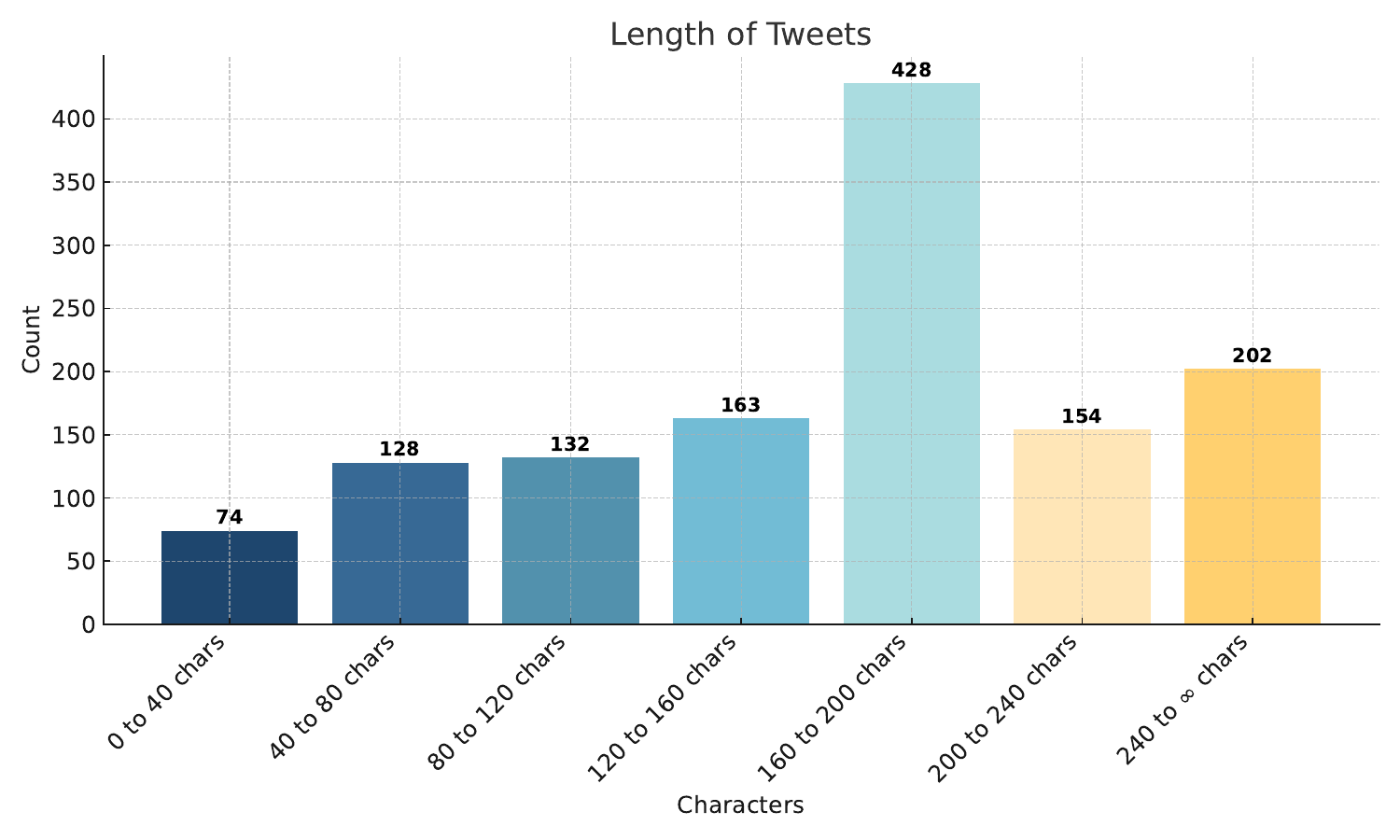}
    \end{minipage}
    \label{fig:length_of_tweets}
    }
    \subfigure[Language distribution]{
    \begin{minipage}[t]{0.3\textwidth}
    \centering
    \includegraphics[width=2.2in]{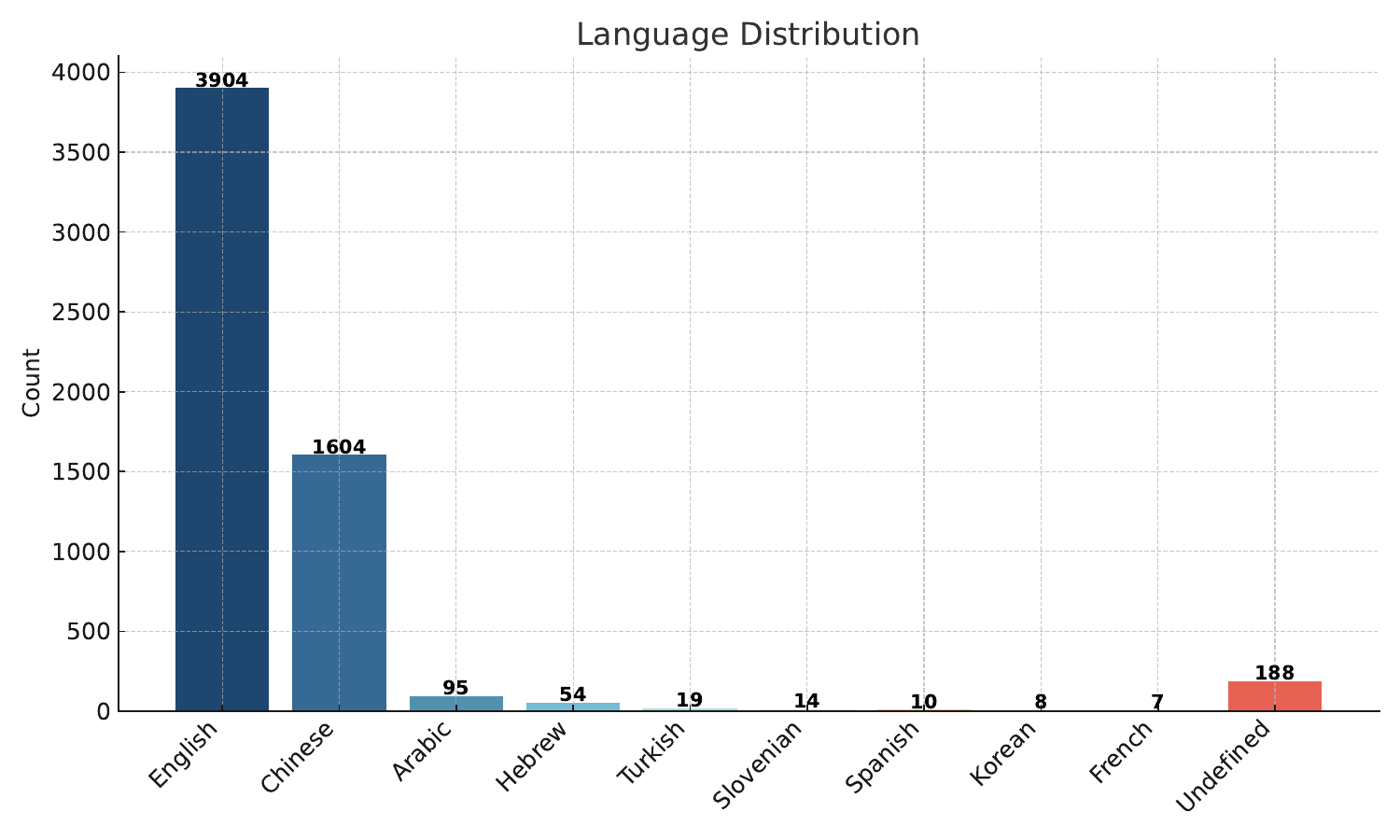}
    \end{minipage}
    \label{fig:language_distribution}
    }
    \subfigure[Contributor influence]{
    \begin{minipage}[t]{0.3\textwidth}
    \centering
    \includegraphics[width=2.2in]{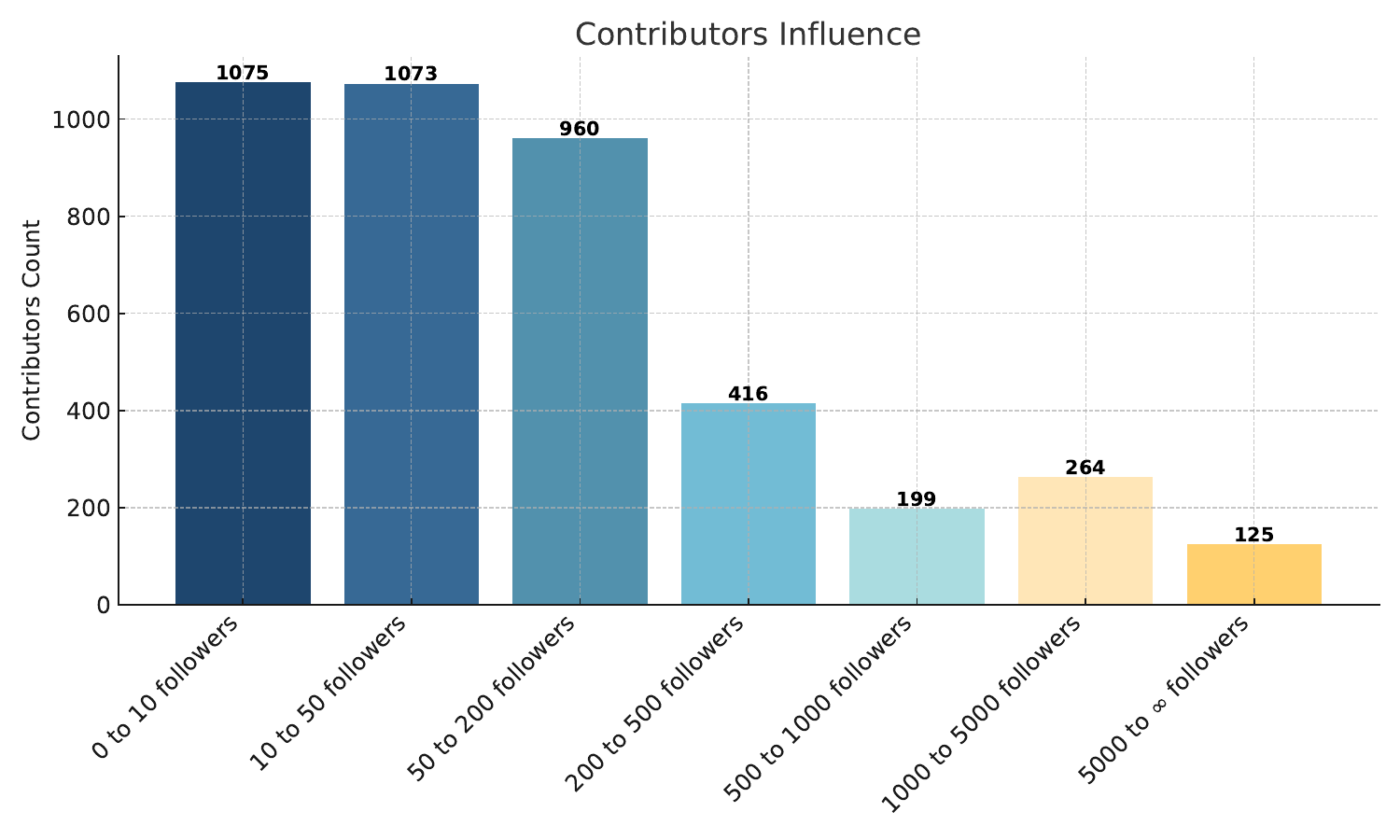}
    \end{minipage}
    \label{fig:contributor_influence}
    }
\caption{Tweet stats related to BRC-20}
\label{fig:tweets-info-1}
\vspace{-0.15in}
\end{figure*}

\subsubsection{\textbf{Tweet stats}}
We conduct an examination of Twitter data surrounding BRC-20. Notably, most contributors opt for web-based tweeting (Fig.\ref{fig:source_distribution}), indicating a higher level of attention spent when accessing BRC-20 content compared to mobile users. Furthermore, the distribution of tweet data is well-balanced (Fig.\ref{fig:tweets_per_contributor}), supported by the fact that the majority of contributors post just one tweet. This minimizes the potential for biased outcomes stemming from excessive tweeting by a single individual.

We have knowledge that a predominantly neutral sentiment observed across users, tweets, and impact suggests a cautiously optimistic view within the community (Fig.\ref{fig:senti-tweets}). This relative optimism is reinforced by the fact that the majority of tweets are longer (160 to 200 characters, Fig.\ref{fig:length_of_tweets}).

The diversity in the age of Twitter accounts engaged in the conversation, ranging from newly created to those over six years old, reveals an appeal that transcends different segments of the community (Fig.\ref{fig:age_of_accounts}). The broad international interest, as evidenced by the primary languages being English and Chinese, underlines the global appeal of BRC-20 (Fig.\ref{fig:language_distribution}).

In terms of influence, the participation across various follower counts, from micro-influencers to major influencers, highlights an inclusive conversation that extends beyond a niche audience (Fig.\ref{fig:contributor_influence}). The consistency in engagement, regardless of the number of followers, adds credibility to the BRC-20 conversation.

\begin{figure}[!hbt]
    \centering
    \includegraphics[width=0.8\columnwidth]{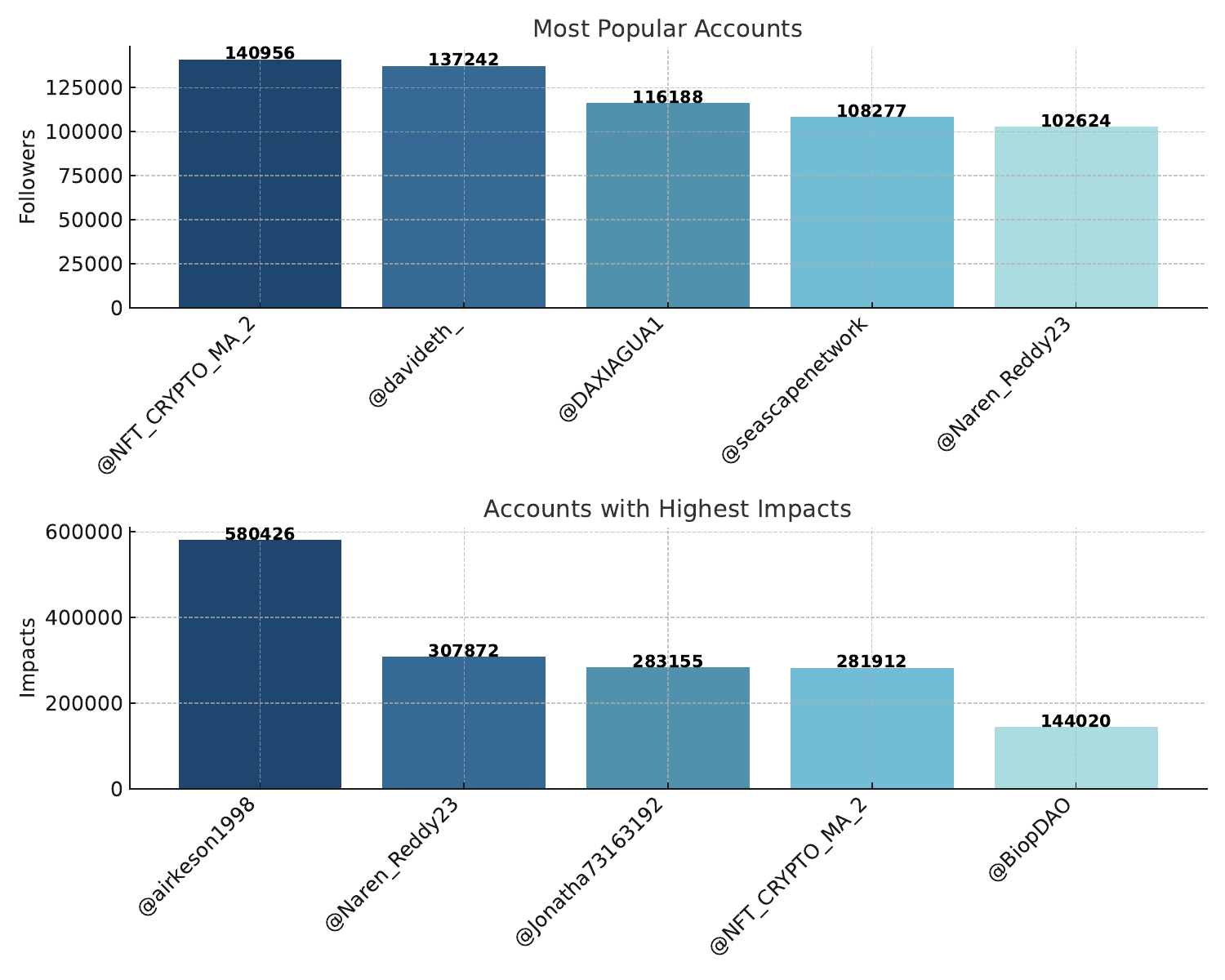}
    \caption{Popular users with the highest impact.}
    \label{fig:tweets-info-2}
    \vspace{-0.15in}
\end{figure}

\smallskip
\noindent \colorbox{pink!50}{\textcolor{black}{\textbf{\text{Finding-V.\ding{203}:}}}} \textit{\textbf{
BRC-20 appeals to users across various regions and age groups.}}

\smallskip
\subsubsection{\textbf{(Non-)Scam among users}}
We first analyze the relationship between user types (normal users vs influencers) and tweet types (scam vs non-scam) in the context of the BRC-20 hashtag. Users were categorized based on specific criteria: influencers were identified from the ``Most Popular'' and ``Highest Impact'' lists, while normal users were those not listed as influencers. Tweets were classified as scams or non-scam based on the presence of certain keywords, repeated messages, and patterns indicative of pyramid selling.

\begin{figure}[!hbt]
    \centering
    \includegraphics[width=0.9\columnwidth]{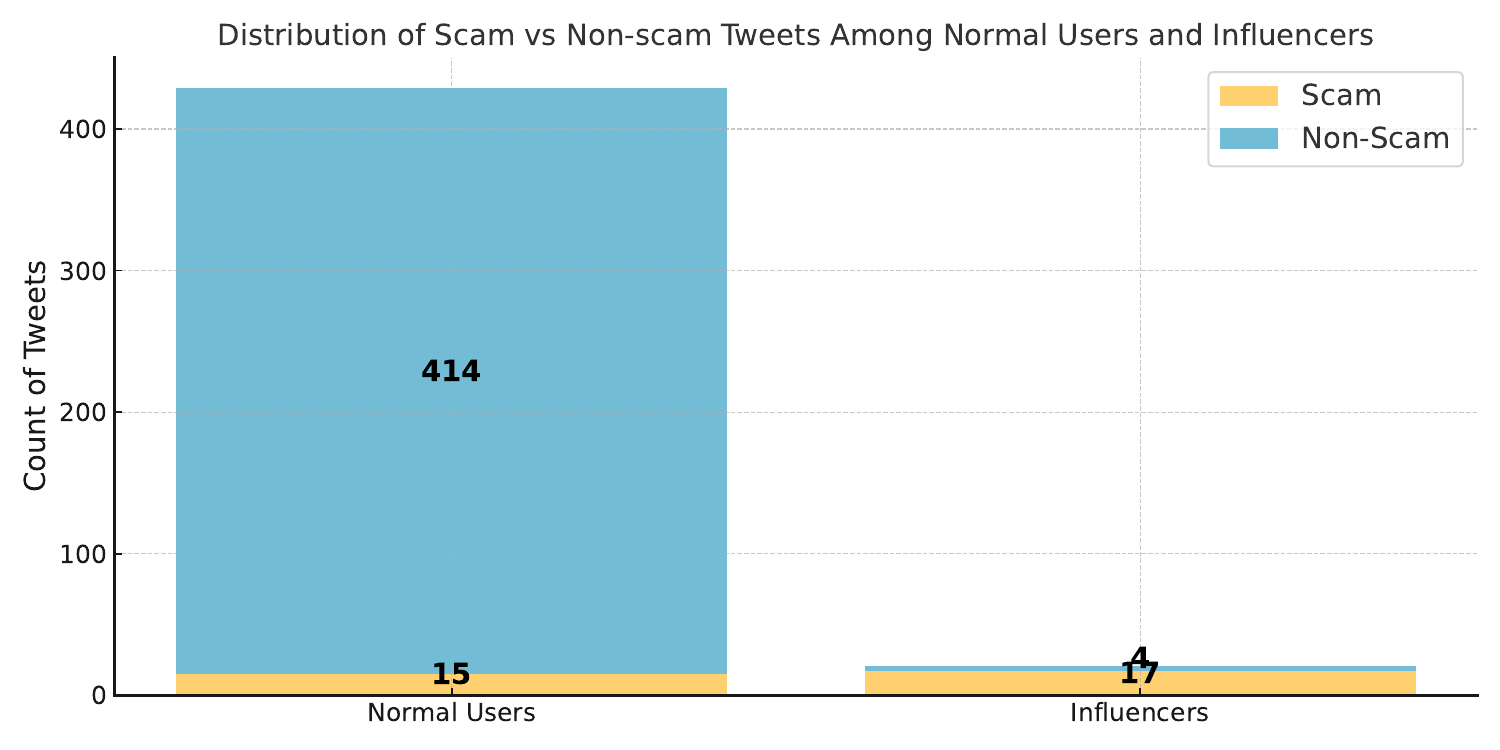}
    \caption{(Non-)Scam tweets among users}
    \label{fig:scam}
\end{figure}

Fig.\ref{fig:scam} unveils a significant distinction between influencers and normal users. While influencers posted fewer tweets overall, a higher proportion of their tweets were classified as scams. This suggests that many influencers may be leveraging their popularity to engage in questionable practices like pyramid selling, possibly with the intent to manipulate the market or deceive followers. The content of their tweets may not reflect a genuine interest in BRC-20, indicating a potential agenda to exploit the hype surrounding the cryptocurrency.

In contrast, normal users predominantly engaged in non-scam tweets, contributing to informative and meaningful discussions about BRC-20. Their engagement pattern reflects a genuine interest in the subject, possibly even involving actual exchange processes of BRC-20. The higher volume of non-scam tweets among normal users reflects authentic interests in BRC-20, unlike the controlled narrative pushed by influencers.

\smallskip
\noindent \colorbox{pink!50}{\textcolor{black}{\textbf{\text{Finding-V.\ding{204}:}}}} \textit{\textbf{While BRC-20 holds the risk of artificial manipulation, the dominantly controlling influence remains within legal and constructive boundaries.}}

\section{Investigation Compared to Historical Peaks}
\label{sec-historical}

\noindent\textbf{Investigation overview.} We conduct an examination of historical crypto market data spanning ten years from 2013 to 2023, encompassing nine prominent tokens including BTC, LTC, DOGE (BRC-type), ETH, BNB, AVA (ERC-type), USDT, USDC, and BUSD (stablecoin). By correlating this historical data with major real-world market waves, we aim to discern if the peaks or prosperity of each market coincide with significant narratives. This macroscopic analysis provides insights into whether BRC represents a genuine wave in tokenomics.

\subsection{Tokenwaves in History}

Based on these price trends, several notable waves in the token market are obtained. The initial peak, predating 2013, can be attributed to the flourishing crypto market driven primarily by the fervor surrounding Bitcoin mining activities and its PoW mechanism~\cite{bonneau2015sok}. As a pioneering force, Bitcoin's impact set the stage for the valuation of the entire cryptocurrency landscape. Following this, the subsequent peak around 2017 aligns with Ethereum's development, sparking a surge in \textit{initial coin offerings} (ICOs)~\cite{bellavitis2021comprehensive}. ICOs facilitated fund-raising by exchanging Ethereum (ETH) via the ERC20 standard~\cite{erc20} for native tokens of various projects, thereby attracting widespread user engagement and diverse investments. This wave was later succeeded by \textit{initial exchange offerings} (IEOs)~\cite{ieo} and analogous \textit{initial development offerings} (IDOs)~\cite{ido}.

Following a two-year cooling-off period, a notable resurgence took place in the mid-2020s, characterized by the rise of \textit{decentralized finance} (DeFi)~\cite{werner2022sok}. DeFi encompasses a range of on-chain financial protocols that mirror traditional market functions, including lending, borrowing, contracts, leverage, and securities. Subsequently, starting in 2021, the spotlight shifted to \textit{non-fungible tokens} (NFTs)\cite{wang2021non} within the Ethereum ecosystem. These distinct digital assets are utilized to represent ownership or validate authenticity for digital artworks, collectibles, and virtual real estate. This trend was further propelled by subsequent developments like the \textit{play-to-earn} concept\cite{yu2022sok} and the growing influence of \textit{Web3}~\cite{wang2022exploring} in 2022. As we progress into 2023, the continued activity in the token space remains evident with the deployment, minting, and transfer of inscriptions on the Bitcoin network via \textit{BRC-20}~\cite{binancebrc}.

\begin{figure}[!hbt]
    \centering
    \includegraphics[width=0.95\columnwidth]{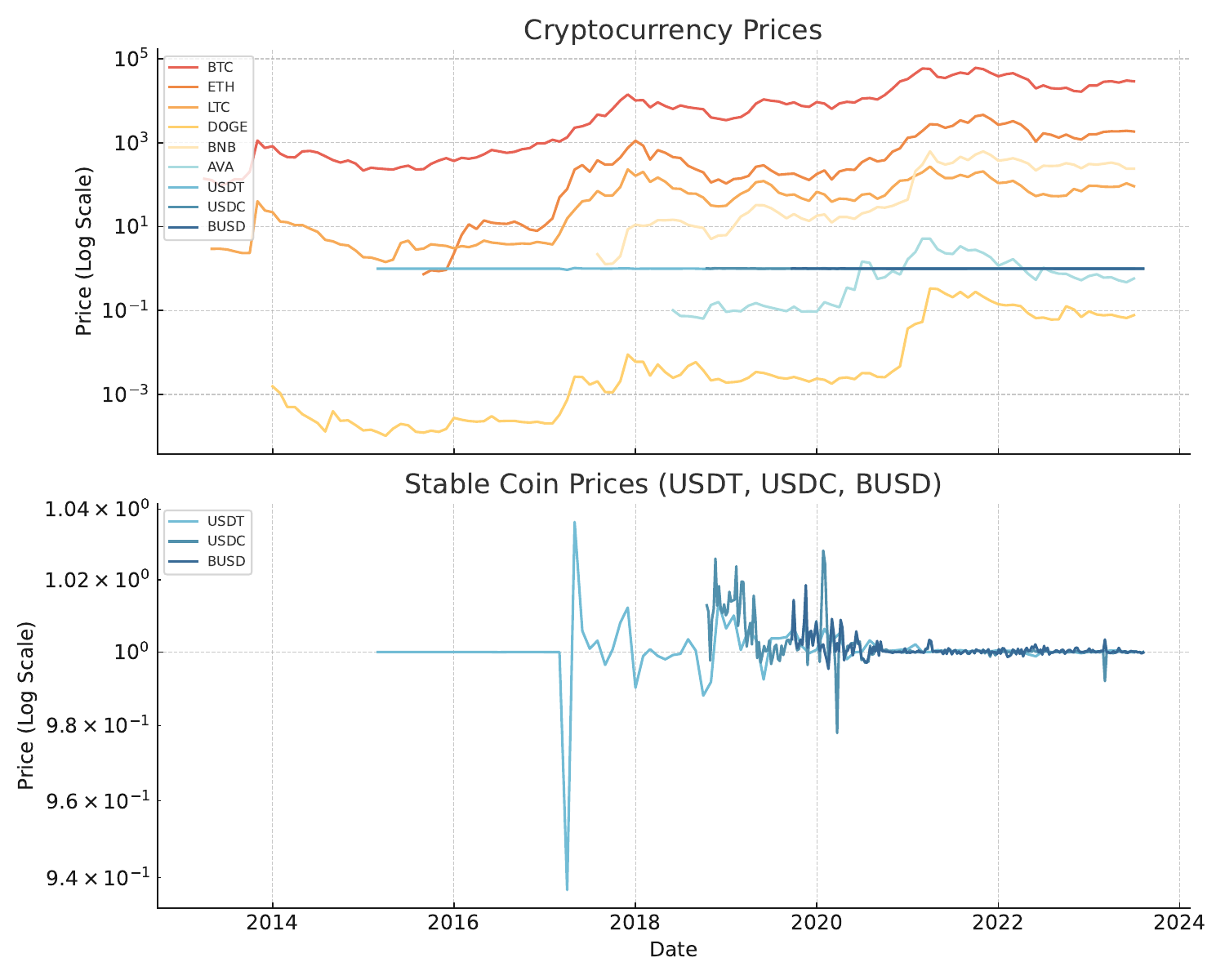}
    \caption{Cryptocurrency prices all through years}
    \label{fig:2013_2023_prices}
\end{figure}

\smallskip
\noindent \colorbox{pink!50}{\textcolor{black}{\textbf{\text{Finding-VI.\ding{202}:}}}} \textit{\textbf{BRC-20 appears to be emerging as a new narrative during 2023, propelling a fresh tokenwave.}}

\subsection{Comparison and Correlation} 

We observed a common movement pattern among most tokens (both BRC-like and ERC-like) except for stablecoins (including USDT, USDC, BUSD). This suggests that token prices are intrinsically interconnected and are influenced by dominant tokens like BTC and ETH. Stablecoins, on the other hand, exhibit a distinct trend, remaining independent of market tokens and maintaining stable values pegged to the US dollar. The broader wave of tokenomics appears to have minimal impact on their fundamental value, except in cases like the Luna-UST collapse \cite{fu2022ftx} where major design flaws were evident. 
We can infer that the surge in Bitcoin prices during the BRC's popularity period indirectly amplifies positive sentiments across the entire token market.

\smallskip
\noindent \colorbox{pink!50}{\textcolor{black}{\textbf{\text{Finding-VI.\ding{203}:}}}} \textit{\textbf{
The patterns of BRC-20 waves align with broader trends observed in the cryptocurrency market.}}

\section{Investigation From Inherent Features}
\label{sec-discussion}

\noindent\textbf{Investigation overview.} In contrast to previous quantitative measurements, this section presents a qualitative evaluation from three perspectives: a comparison with other standards, its positive attributes and impacts, as well as notable limitations that must not be disregarded.

\vspace{-0.05in}
\subsection{Compare with Existing Standards} \label{subsec-quali-compri}

The majority of token standards in competitive blockchains (summarised in \underline{Tab.\ref{tab:erc}}), such as BEP-20/721 (BNB Smart Chain), ARC-20/721 (Avalanche), and XRC-20/721 (XDC Network \cite{xrc20}), draw inspiration from the Ethereum repository. These ERC-like standards share common attributes, adhering to the 20-track standard for fungibility and the 721-track standard for non-fungibility. NFTs in these chains possess programmable smart contracts, allowing for limitless issuance.

\begin{table}[!hbt]
\caption{Comparison with competitive standards}\label{tab:erc}
\renewcommand\arraystretch{1.1}
\begin{center}
\resizebox{\linewidth}{!}{
\begin{tabular}{l|ccc|ccc|c} 
        \textbf{Standard} & \textbf{Time} & \textbf{Network} &\rotatebox{90}{\textbf{Upper-bound}}  &  \rotatebox{90}{\textbf{Fungibility}} &   \rotatebox{90}{\textbf{Divisibility} }  &  \rotatebox{90}{\textbf{Transferrability}} & \textbf{Application} \\
        \midrule
        \cellcolor{gray!10}ERC-20 & 2015 & Ethereum & \cmark  & \cmark & \cmark &  \cmark (Tx)  & Currency \\ 
        \cellcolor{gray!10}ERC-721 & 2017 & Ethereum & \xmark  & \xmark &  \xmark  & \cmark (SC)  & NFT\\ 
        \cellcolor{gray!10}ERC-1155 & 2018 & Ethereum &\xmark  & semi &    \xmark   & \cmark (SC)  & Game \\ 
        \cellcolor{gray!10}ERC-3525 & 2022 & Ethereum & \xmark & semi &   \cmark   & \cmark (SC)& Equity \\ 
        \cellcolor{gray!10}ERC-3475 & 2022 & Ethereum &\xmark  & semi &   n/a &   \cmark (SC) & Equity \\ 
        \midrule
        \cellcolor{gray!10}BEP-20 & 2021 & BSC & \cmark  & \cmark &  \cmark  & \cmark (Tx)  & Currency \\ 
        \cellcolor{gray!10}BEP-721 & 2022 & BSC & \xmark  & \xmark &   \xmark  & \cmark (SC)  & NFT \\ 
        \cellcolor{gray!10}ARC-20 & 2022 & Avalanche & \cmark & \cmark &    \cmark  & \cmark (Tx)  & Currency \\
        \cellcolor{gray!10}ARC-721 & 2022 &  Avalanche & \xmark & \xmark &   \xmark   & \cmark (SC) & NFT\\
        \cellcolor{gray!10}XRC-20 & 2023 & XDC & \cmark  & \cmark &  \cmark  & \cmark (Tx)  & Currency \\ 
        \cellcolor{gray!10}XRC-721 & 2023 & XDC & \xmark  & \xmark &   \xmark  & \cmark (SC)  & NFT \\ 
        \midrule
        \cellcolor{gray!10}DRC-20 & 2023 & DogeCoin & \cmark  & \xmark &   \xmark  & \cmark (Tx)  & NFT \\  
        \cellcolor{gray!10}LTC-20 & 2023 & Litecoin &  \cmark & \xmark &  \xmark  & \cmark (Tx)  & NFT \\ 
        \cellcolor{gray!10}\textbf{BRC-20} & \cellcolor{gray!10}2023 &  \cellcolor{gray!10}Bitcoin & \cellcolor{gray!10}\cmark   & \cellcolor{gray!10}\xmark & \cellcolor{gray!10}\xmark  & \cellcolor{gray!10}\cmark (Tx)  & \cellcolor{gray!10}NFT \\ 
        \bottomrule
\end{tabular}
}
\end{center}
\vspace{-0.2in}
\end{table}

Contrastingly, BRC-like standards~\cite{drc20}\cite{lrc20} integrate uniqueness into transaction payloads, stemming from their limited units (sats). This results in non-fungible tokens being transacted through a combination of regular transactions and specific operations. On the flip side, ERC-like standards achieve distinctiveness via a parameter called the token ID in functions (cf. Algm.\ref{alg-erc}), potentially utilizing various functions in an upper-layer operation. This gives rise to diverse token standards with features like 1155/3525/3475. Transfers within this framework rely on state transitions facilitated by contracts operating on the chain. We present more differences in \underline{Tab.\ref{tab:nfts}}.

This divergence also translates into disparities in popularity. ERC-compatible chains thrive with active developers and Dapps, attracting a larger user base. Conversely, BRC-like chains often grapple with a dearth of active developers, hampering the initiation of innovative approaches.

\vspace{-0.1in}
\begin{algorithm} 
\caption{\textbf{ERC-721 Standard Interfaces}}\label{alg-erc}
\BlankLine
 \textbf{interface} ERC721 \{ \\
 \quad    function \textcolor{teal}{$\mathsf{ownerOf}$}(uint256$\_$\textcolor{magenta}{tokenID}) external view returns (address); \\
 \quad    function \textcolor{teal}{$\mathsf{transferFrom}$}(address$\_$from, address$\_$to, uint256$\_$\textcolor{magenta}{tokenId}) external payable;  ... \} \\    
\label{algo:rnft}
\end{algorithm}
\vspace{-0.1in}

\smallskip
\noindent \colorbox{pink!50}{\textcolor{black}{\textbf{\text{Finding-VII.\ding{202}:}}}} \textit{\textbf{BRC-20 stands out distinctly from ERC-like standards due to its structure, leading to a shortage of active developers and on-chain applications.}}

\subsection{Advantages To be Highlighted}\label{subsec-quali-advan}

\smallskip
\noindent\textbf{\text{System stability.}} Stability is primarily dependent on the network of distributed miners and their commitment. Augmented stability is achieved through two primary avenues.

\begin{itemize}
  \item \textit{New players.} As explained in Sec.\ref{sec-construction}, the tracing of each satoshi requires the utilization of ORD software. This means that despite the availability of user-centric solutions like Ordinals markets, individuals wanting full control over the entire Ordinal procedure and the creation of an Inscription must operate a Bitcoin full node (rather than a lightweight node). This element, among others, has led to a marked rise in accessible Bitcoin nodes. The more active Bitcoin full nodes there are, the greater the decentralization of the Bitcoin network becomes.

  \item \textit{Increased revenue.} The incorporation of ordinal inscriptions intensifies congestion within the Bitcoin blockchain, leading to an upward trajectory in fees and bolstering miners' earnings. This provides miners with advantages and enhances their commitment to the system. This advancement holds promise for the long-term sustainability of the Bitcoin blockchain, as its viability heavily relies on substantial transaction fees. The introduction of supplementary application layers, like ordinals, holds the potential to sustain heightened congestion. This, in turn, alleviates concerns about liquidity shortages or inadequate transaction volumes.

\end{itemize}

\noindent\textbf{\text{Infrastructure construction.}} Driven by BRC, Bitcoin's advancements in infrastructure and DApps have also led to substantial progress (also refer to \underline{Tab.\ref{tab:tool}}). Notably, Bitcoin wallets like Hiro and Xverse have rapidly expanded their support for BRC-related protocols, swiftly introducing products such as the BRC Explorer. Additionally, even the Bitcoin NFT market, traditionally centered around Stacks-based projects, has undergone a transformation with the recent launch of Gamma's Ordinals marketplace. Following closely, Magic Eden introduced its Bitcoin NFT marketplace. Esteemed NFT studios such as Yuga Labs and DeGods have also joined this movement, unveiling Ordinals-based projects within the past month. This surge in innovation is not confined to Bitcoin's base layer; it's equally evident within Bitcoin Layer2 solutions like the Lightning Network, Liquid, Rootstock, and Stacks.

\smallskip
\noindent \colorbox{pink!50}{\textcolor{black}{\textbf{\text{Finding-VII.\ding{203}:}}}} \textit{\textbf{The emergence of BRC enhances system stability and fosters the development of complementary tools.}}

\subsection{Limitations Cannot Be Ignored}\label{subsec-quali-disadv}

\noindent\textbf{\text{Costly.}}  We noted that the protocol requires two on-chain transactions to complete the transfer operation, which is costly and less user-friendly, while additionally, the granularity of exchanges is limited to the amounts defined in each PSBT. Moreover, the inscribed satoshis become invalid after use, necessitating the inscription of a new satoshi for each new transaction, which deviates from the original concept of Ordinals as long-lasting, meaningful inscriptions. 

\smallskip
\noindent\textbf{\text{Increased fees.}} Similarly, analyzing transaction fees spanning December 2022 to April 2023, as outlined in \cite{bertucci2023bitcoin}, it's evident that significant total fees accrue with substantial inscriptions, signifying larger transactions. Importantly, a clear positive correlation emerges between Bitcoin ordinal inscriptions and transaction fees across diverse transactions, which contributes to the overall block fees. Consequently, the integration of ordinal inscriptions amplifies congestion within the Bitcoin blockchain, resulting in an upward trajectory of fees. This will raise concerns for regular users.

\smallskip
\noindent\textbf{\text{Stateless.}} BRC-20 and Ordinals continue to grapple with the challenge posed by the inherently stateless nature of UTXO transactions and the state models that most applications demand. The advancement of these protocols and their capacity to accommodate more comprehensive functionalities, including a versatile virtual machine, hinges on the ongoing market incentives and the sustained appreciation of coin value.

\smallskip
\noindent\textbf{\text{Centralization.}}  The escalating size of the Bitcoin network may discourage users from running their nodes due to increased requirements for downloading a copy of the network. Currently, most BRC-20 wallets necessitate running a full node, a practice not commonly embraced by regular users. As a result, users resort to third-party APIs, potentially creating centralized security vulnerabilities. Although different indexers can be connected for cross-validation, this requires additional steps and understanding from the users' side.

\smallskip
\noindent\textbf{\text{Meme-nature.}} Presently, a significant portion of the BRC-20 tokens in circulation, such as ORDI, PEPE, MOON, and others, predominantly belong to the category of meme coins. Due to the absence of consensus among communities and the lack of support for smart contracts, these tokens offer minimal practical utility and are notably swayed by trends in social media sentiment. Although this phenomenon sparks speculative interest, the tokens' limited functionality and the consequent dearth of a robust holder base suggest a potential vulnerability to abrupt, unforeseen value declines.

\smallskip
\noindent \colorbox{pink!50}{\textcolor{black}{\textbf{\text{Finding-VII.\ding{204}:}}}} \textit{\textbf{BRC brings network congestion, leading to increased reliance on centralized tools and rising fees. Additionally, it retains its inherent limitation of extensibility.}}

\section{User Perception and Implications}
\label{sec-conclusion}

\subsection{Reality and Misconceptions in User Perception} \label{subsec-percep-reality}

\noindent\textbf{Realities.}  Based on aforementioned investigations, several realities emerge from user perceptions in the BRC-20 landscape.

\begin{itemize}

\item \textit{Genuine interest in BRC-20.} Users exhibit an enthusiastic interest in novel crypto concepts, actively participating in discussions (\textbf{V.}\ding{203}) across social media platforms (\textbf{V.}\ding{202}). They demonstrate their commitment to the market by investing time and funds, which is reflected by its market performance (\textbf{IV.}\ding{202}) and a trend of tokenwaves (\textbf{VI.}\ding{202}).

\item \textit{Noteworthy captical returns.} BRC-20 tokens present a remarkable market performance, showcasing substantial returns (\textbf{IV.}\ding{204}) that outpace the performance of equivalent tokens in other categories (\textbf{IV.}\ding{204}\&\ding{206}).

\item \textit{Interconnected ecosystem.} The BRC-20 ecosystem reveals an interconnected network of tokens (\textbf{IV.}\ding{207}), indicating a close interdependence of user perceptions and behaviors within this specific subset of tokens.

\item \textit{Driving innovation.} Moreover, the advent of BRC-20 has acted as a catalyst for driving innovation in the field, leading to the development of complementary tools and contributing to the overall stability of the system (\textbf{VII.}\ding{203}).

\end{itemize}

\noindent\textbf{Misconceptions.} In contrast, our investigations have also uncovered certain misconceptions.

\begin{itemize}

\item \textit{Ephemeral enthusiasm.} User enthusiasm for new concepts often follows a cyclical pattern of initial excitement followed by a potential decline in engagement (\textbf{IV.}\ding{202}), particularly if immediate benefits are not realized  (\textbf{IV.}\ding{205}\&\ding{206}).

\item \textit{Limited market size.} BRC-related markets still occupy a relatively small share compared to larger markets like Bitcoin's daily transaction volume (\textbf{IV.}\ding{208}) or the market capitalization of ERC-like tokens (\textbf{IV.}\ding{203}).

\item \textit{Dependency on dominance.} Much like derivative tokens, the trend of many BRC-20 tokens appears to be influenced by a select few dominant projects such as Ordi (\textbf{VI.}\ding{203}), as well as social influencers (\textbf{V.}\ding{204}).

\item \textit{One-Sided development.} The majority of developed tools are built upon existing data sources like web browsers or account-related APIs, rather than introducing novel logical innovations like those found in smart contracts—reflecting an inherent limitation (\textbf{VII.}\ding{202}\&\ding{204}).

\end{itemize}



\begin{table}[!hbt]
\caption{NFT Comparisons}\label{tab:nfts}
\resizebox{\linewidth}{!}{
\begin{tabular}{rll} 
\toprule

& \textbf{Bitcoin NFT} & \textbf{Other NFTs} \\ 
\midrule
\textbf{Protocol form} & \cellcolor{gray!10}Ordinal & ERC-721, ERC-1155, SPL \\ 
\textbf{Description} & \cellcolor{gray!10}Inscription & NFT \\
\textbf{Storage} & \cellcolor{gray!10}Entirely on-chain & Partially on IPFS/Arweave \\ 
\textbf{Code update} & \cellcolor{gray!10}Not allowed & Depends on contract code \\ 
\cmidrule{1-1}
\textbf{Minting} & \cellcolor{gray!10}\makecell[l]{Not possible without a node, need \\ via third-party designed services} & \makecell[l]{DMostly can directly interact\\ with the webpage} \\ [1ex] 
\textbf{Trading} & \cellcolor{gray!10}NFT marketplace & NFT Marketplace \\ 
\cmidrule{1-1}
\textbf{Extensibility} & \cellcolor{gray!10}\makecell[l]{Difficult due to Bitcoin's \\  scripting limitations} &  \makecell[l]{Easier due to programmable \\  smart contracts } \\ 
\textbf{Consumption} & \cellcolor{gray!10}High due to PoW consensus & Up to platform consensus \\ 
\cmidrule{1-1}
\textbf{Pros} & \cellcolor{gray!10}Scarcity, rarity-nature & \makecell[l]{Mainstream contract mode, \\ high user base} \\ 
\textbf{Cons} & \cellcolor{gray!10}\makecell[l]{Low block speed, no bulk minting; \\ Difficulties in minting/trading; \\ Wallet entry is complex} &  \makecell[l]{No special gimmicks or fame,\\ easily overlooked} \\ 

\bottomrule

\end{tabular}
}
\end{table}

\subsection{Towards Enhancement} \label{subsec-percep-enhance}

\noindent\textbf{Improving user awareness by education.} As our investigations revealed,  a prevalent lack of understanding among both non-professional and professional users regarding the fundamental concepts of BRC, and even the operational intricacies of Bitcoin itself, let alone the workings of BRC within the Bitcoin network. This limited comprehension leads to sparse discussions in public channels, with mere hundreds of tweets about BRC compared to thousands about Ethereum or even millions about a popular singer's new song. Among these mentions, the majority remain superficial, lacking substantive content. To enhance users' awareness and understanding of BRC and Bitcoin NFTs, two viable approaches stand out. Firstly, the establishment of an educated community through platforms like MOOCs and easily accessible YouTube videos could be pivotal. Open forums could address security concerns, while independent implementations on GitHub channels could offer potential solutions. For instance, BRC has been interpreted by prominent companies and media outlets, like Binance. Secondly, facilitating the creation of competing BRC services and third-party tools by developers can yield quick responses to user needs, encompassing NFT-related functions such as creation, purchase, auction, and exchange, particularly for technical users. Several third-party tools have already emerged for BRC to aid users in facilitating user experiences.

\smallskip
\noindent\textbf{Encouraging communities for further engagement.}  Independent tools and services have consistently occupied a prominent position within the space of BRC-20 and its associated communities. Diverse applications have been developed to enhance the user experience of BRC-based products. For instance, volunteers from Cryptokoryo \cite{duneanalytic} and Dataalways \cite{duneanalytic1} have created statistical services that showcase insightful trends related to Ordinals, Inscriptions, and BRC-tokens. Additionally, various media outlets provide dedicated sections to succinctly summarize the latest news relevent to BRC. BRC explorers have also been implemented to provide real-time price fluctuations. These tools significantly contribute to increasing user understanding of basic mechanisms while alleviating concerns about potential drawbacks. The seamless integration of third-party tools with other existing services, in particular DeFi protocols \cite{jiang2023decentralized} and cross-chain technologies \cite{wang2023exploring}, adds value and has the potential to enhance adoption.

\smallskip
\noindent\textbf{Attracting new attentions.} BRC-20 also draws inspiration from the NFT landscape, which has demonstrated remarkable growth over the past couple of years. Users who have actively engaged in NFT trading and gaming activities (such as minting, participating in airdrops, etc.) are likely to exhibit an inherent curiosity in exploring BRC NFTs, provided there are no significant barriers. It would be prudent for BRC developers to offer tools that facilitate interoperability across various blockchain ecosystems, including Ethereum, Polygon, Binance Smart Chain, and Avalanche. When compared to new users entering the traditional market, those migrating from established Web3 ecosystems yet to be fully developed offer a vast and readily accessible user base.

\section{Conclusion}
\label{sec-conclusion}

In this paper, we dig into the novel concept of BRC-20. We elucidate its operational mechanisms and empirically conduct a range of tangible investigations, encompassing market performance and user sentiments. Recognizing that user perception plays a pivotal role in shaping the nature of BRC, we subsequently explore the dichotomy between hope and hype, which significantly influence user perception. Our findings lead to a conservative conclusion that while BRC-20 represents a promising inception within the Bitcoin ecosystem, it may not attain the same level as ERC-like ecosystems.

\normalem
{\footnotesize \bibliographystyle{IEEEtran}
\bibliography{bib}}

\begin{thebibliography}{10}
\providecommand{\url}[1]{#1}
\csname url@samestyle\endcsname
\providecommand{\newblock}{\relax}
\providecommand{\bibinfo}[2]{#2}
\providecommand{\BIBentrySTDinterwordspacing}{\spaceskip=0pt\relax}
\providecommand{\BIBentryALTinterwordstretchfactor}{4}
\providecommand{\BIBentryALTinterwordspacing}{\spaceskip=\fontdimen2\font plus
\BIBentryALTinterwordstretchfactor\fontdimen3\font minus
  \fontdimen4\font\relax}
\providecommand{\BIBforeignlanguage}[2]{{%
\expandafter\ifx\csname l@#1\endcsname\relax
\typeout{** WARNING: IEEEtran.bst: No hyphenation pattern has been}%
\typeout{** loaded for the language `#1'. Using the pattern for}%
\typeout{** the default language instead.}%
\else
\language=\csname l@#1\endcsname
\fi
#2}}
\providecommand{\BIBdecl}{\relax}
\BIBdecl

\bibitem{nakamoto2019bitcoin}
S.~Nakamoto, ``Bitcoin: A peer-to-peer electronic cash system,'' Manubot, Tech.
  Rep., 2019.

\bibitem{wood2014ethereum}
G.~Wood \emph{et~al.}, ``Ethereum: A secure decentralised generalised
  transaction ledger,'' \emph{Yellow paper}, vol. 151, no. 2014, pp. 1--32,
  2014.

\bibitem{delgado2019analysis}
S.~Delgado-Segura, C.~P{\'e}rez-Sola, G.~Navarro-Arribas, and
  J.~Herrera-Joancomart{\'\i}, ``Analysis of the {Bitcoin} {UTXO} set,'' in
  \emph{Financial Cryptography and Data Security Workshops, BITCOIN, VOTING,
  and WTSC (BITCOIN@FC)}.\hskip 1em plus 0.5em minus 0.4em\relax Springer,
  2018, pp. 78--91.

\bibitem{binancebrc}
B.~Research, ``{BRC}-20 tokens: A primer,''
  \emph{\url{https://research.binance.com/static/pdf/BRC-20\%20Tokens\%20-\%20A\%20Primer.pdf}},
  2023.

\bibitem{erc20}
V.~Fabian and B.~Vitalik, ``Ethereum {ERC-20} token standard,''
  \emph{Accessible: \url{https://eips.ethereum.org/EIPS/eip-20}}, 2015.

\bibitem{brc20experiment}
Anonymous, ``{BRC}-20 experiment,''
  \emph{\url{https://domo-2.gitbook.io/brc-20-experiment/}}, 2023.

\bibitem{wang2021non}
Q.~Wang, R.~Li \emph{et~al.}, ``Non-fungible token ({NFT}): Overview,
  evaluation, opportunities and challenges,'' \emph{arXiv preprint
  arXiv:2105.07447}, 2021.

\bibitem{binance1}
B.~Research, ``A new era for {Bitcoin},''
  \emph{\url{https://research.binance.com/static/pdf/a-new-era-for-bitcoin.pdf}},
  2023.

\bibitem{ordinalwallet}
``Ordinal wallet,'' \emph{\url{https://ordinalswallet.com/brc20}}, 2023.

\bibitem{unisat}
``Unisat wallet: Open source chrome extension for {Bitcoin} ordinals \&
  {BRC}-20,'' \emph{\url{https://unisat.io/brc20?ref=bankless.ghost.io}}, 2023.

\bibitem{duneanalytic}
cryptokoryo, ``Dune analytics: {Bitcoin} tokens, {BRC}-20 ({BRC20}
  {Ordinals}),'' \emph{\url{https://dune.com/cryptokoryo/brc20}}, 2023.

\bibitem{duneanalytic1}
dataalways, ``Dune analytics: {Ordinals} - {Inscriptions} on {Bitcoin}),''
  \emph{\url{https://dune.com/dataalways/ordinals}}, 2023.

\bibitem{bipord}
C.~Rodarmor, ``Ordinal numbers,''
  \emph{\url{https://github.com/casey/ord/blob/master/bip.mediawiki}}, 2022.

\bibitem{ordinbook}
``Ordinal theory handbbok,'' \emph{\url{https://docs.ordinals.com/}}, 2023.

\bibitem{binance2}
B.~Research, ``Monthly market insights,''
  \emph{\url{https://research.binance.com/static/pdf/monthly-market-insights-2023-06-.pdf}},
  2023.

\bibitem{bertucci2023bitcoin}
L.~Bertucci, ``Bitcoin ordinals: Determinants and impact on total transaction
  fees,'' \emph{Available at SSRN 4486127}, 2023.

\bibitem{kiraz2023nft}
M.~S. Kiraz, E.~Larraia, and O.~Vaughan, ``{NFT} trades in {Bitcoin} with
  off-chain receipts,'' \emph{Cryptology ePrint Archive}, 2023.

\bibitem{news1}
Ann, ``Are {Ordinals} and {BRC}-20 the future of {Bitcoin}?''
  \emph{\url{https://medium.com/crypto-24-7/are-ordinals-and-brc-20-the-future-of-bitcoin-171b9c5fe799}},
  2023.

\bibitem{news2}
T.~Machines, ``What are {Bitcoin} ordinals?''
  \emph{\url{https://trustmachines.co/learn/what-are-bitcoin-ordinals-inscriptions/}},
  2023.

\bibitem{news3}
Y.~Max, ``5 things you must know about {Bitcoin} {NFT}s,''
  \emph{\url{https://medium.com/coinmonks/bitcoin-nfts-ordinals-5-questions-to-ask-yourself-before-investing-d26ae71c4e31}},
  2023.

\bibitem{news4}
K.~George, ``Bitcoin {NFT}s: What are {Ordinal} {NFT}s and how do you mint
  one?''
  \emph{\url{https://www.coindesk.com/learn/bitcoin-nfts-what-are-ordinal-nfts-and-how-do-you-mint-one/}},
  2023.

\bibitem{pshow}
Trevor.btc, ``Podcast: The ordinal show,''
  \emph{\url{https://twitter.com/TheOrdinalShow}}, 2023.

\bibitem{antonopoulos2017mastering}
A.~M. Antonopoulos, \emph{Mastering {Bitcoin}: Programming the open
  blockchain}.\hskip 1em plus 0.5em minus 0.4em\relax " O'Reilly Media, Inc.",
  2017.

\bibitem{taproof}
C.~Marguerita, ``Bitcoin's taproot upgrade: What you should know,''
  \emph{\url{https://www.investopedia.com/bitcoin-taproot-upgrade-5210039}},
  2022.

\bibitem{segwit}
F.~Jake, ``Segregated witness ({SegWit}): Definition, purpose, how it works,''
  \emph{\url{https://www.investopedia.com/terms/s/segwit-segregated-witness.asp}},
  2022.

\bibitem{raresociety}
``Rare satoshi society,''
  \emph{\url{https://magiceden.io/ordinals/marketplace/rss}}, 2023.

\bibitem{pizza}
``Bitcoin pizza,''
  \emph{\url{https://academy.binance.com/en/glossary/bitcoin-pizza}}, 2023.

\bibitem{orc20}
``About {ORC}-20: Open token standard for {Bitcoin} ordinal and wider use
  cases,'' \emph{\url{https://docs.orc20.org/}}, 2023.

\bibitem{sharpe1998sharpe}
W.~F. Sharpe, ``The sharpe ratio,'' \emph{Streetwise--the Best of the Journal
  of Portfolio Management}, vol.~3, pp. 169--85, 1998.

\bibitem{bonneau2015sok}
J.~Bonneau, A.~Miller \emph{et~al.}, ``{SoK}: Research perspectives and
  challenges for {Bitcoin} and cryptocurrencies,'' in \emph{IEEE Symposium on
  Security and Privacy (SP)}.\hskip 1em plus 0.5em minus 0.4em\relax IEEE,
  2015, pp. 104--121.

\bibitem{bellavitis2021comprehensive}
C.~Bellavitis, C.~Fisch, and J.~Wiklund, ``A comprehensive review of the global
  development of initial coin offerings ({ICO}s) and their regulation,''
  \emph{Journal of Business Venturing Insights}, vol.~15, p. e00213, 2021.

\bibitem{ieo}
``Wiki: Initial exchange offering ({IEO}),''
  \emph{\url{https://www.wikiwand.com/en/Initial_exchange_offering}}, 2023.

\bibitem{ido}
array.io, ``Initial development offering — a new approach to fundraising,''
  \emph{Accessible:
  \url{https://medium.com/array-io/initial-development-offering-a-new-approach-to-fundraising-fa16f9834f00}},
  2018.

\bibitem{werner2022sok}
S.~Werner, D.~Perez, L.~Gudgeon \emph{et~al.}, ``{SoK}: Decentralized finance
  ({{DeFi}}),'' in \emph{Proceedings of the ACM Conference on Advances in
  Financial Technologies (AFT)}, 2022, pp. 30--46.

\bibitem{yu2022sok}
J.~Yu, M.~Zhang, X.~Chen, and Z.~Fang, ``{SoK}: Play-to-earn projects,''
  \emph{arXiv preprint arXiv:2211.01000}, 2022.

\bibitem{wang2022exploring}
Q.~Wang, R.~Li \emph{et~al.}, ``Exploring {Web3} from the view of blockchain,''
  \emph{arXiv preprint arXiv:2206.08821}, 2022.

\bibitem{fu2022ftx}
S.~Fu \emph{et~al.}, ``{FTX} collapse: a {Ponzi} story,'' \emph{arXiv preprint
  arXiv:2212.09436}, 2022.

\bibitem{xrc20}
``{XRC}-20: The {XDC} network,''
  \emph{\url{https://docs.xdc.org/introduction/standards/xrc20}}, 2023.

\bibitem{drc20}
``Doge labs: {DRC}-20 {Doginals},'' \emph{\url{https://drc-20.org/}}, 2023.

\bibitem{lrc20}
``Litecoin community introduces {LTC}-20: An experimental standard for
  {NFTs},''
  \emph{\url{https://beincrypto.com/litecoin-community-ltc-20-experimental-standard-litecoin-nfts/}},
  2023.

\bibitem{jiang2023decentralized}
E.~Jiang \emph{et~al.}, ``Decentralized finance ({DeFi}): A survey,''
  \emph{arXiv preprint arXiv:2308.05282}, 2023.

\bibitem{wang2023exploring}
G.~Wang \emph{et~al.}, ``Exploring blockchains interoperability: A systematic
  survey,'' \emph{ACM Computing Surveys (CSUR)}, 2023.

\end{thebibliography}

\end{document}